\documentclass[fleqn,usenatbib]{mnras}
\usepackage[utf8]{inputenc}
\usepackage{babel}
\usepackage[T1]{fontenc}
\usepackage{graphicx}
\usepackage{lipsum}
\usepackage[font=small,labelfont=bf,tableposition=top]{caption}
\usepackage{booktabs}
\usepackage{threeparttable}
\usepackage{soul}
\usepackage{amsmath}
\usepackage{newtxtext,newtxmath}

\usepackage{ae,aecompl}
\usepackage{threeparttable}

\usepackage{dblfloatfix}
\usepackage{amsmath}	
\usepackage{threeparttable}
\usepackage{subcaption}
\usepackage{caption}
\usepackage{color}
\usepackage[dvipsnames]{xcolor}
\usepackage[normalem]{ulem}
\usepackage{booktabs}
\usepackage{graphicx}

\newcommand{\fermi}{\textit{Fermi}-{\rm LAT}}

\newcommand{\chandra}{\textit{Chandra}}

\newcommand{\gray}{$\gamma$-ray~}

\newcommand{\xray}{X-ray~}
\newcommand{\xrays}{X-rays~}

\newcommand{\msun}{\mbox{$\rm M_\odot$}}

\newcommand{\nh}{\mbox{$N_{\rm H}$}}

\def\deg{\hbox{$^\circ$}}

\title[Shear acceleration in kpc-scale FR II jets]{Studying \xray spectra from large-scale jets of FR II radio galaxies: application of shear particle acceleration}
\author[He et al.]{
Jia-Chun He$^{1}$, 
Xiao-Na Sun$^{1}$\thanks{xiaonasun@gxu.edu.cn}, 
Jie-Shuang Wang$^{2}$\thanks{jswang@mpi-hd.mpg.de}, 
Frank M. Rieger$^{3,2}$
Ruo-Yu Liu$^{4,5}$, 
\newauthor
En-Wei Liang$^{1}$\\ 
$^{1}$Guangxi Key Laboratory for Relativistic Astrophysics, School of Physics Science and Technology, Guangxi University, Nanning 530004, China\\
$^{2}$Max-Planck-Institut f{\"u}r Kernphysik, P.O. Box 103980, 69029 Heidelberg, Germany\\
$^{3}$Institute for Theoretical Physics, University of Heidelberg, Philosophenweg 12, D-69120 Heidelberg, Germany\\
$^{4}$School of Astronomy and Space Science, Nanjing University, Nanjing 210093, China\\
$^{5}$Key Laboratory of Modern Astronomy and Astrophysics (Nanjing University), Ministry of Education, Nanjing 210023, China
}

\begin{document}
\label{firstpage}
\pagerange{\pageref{firstpage}--\pageref{lastpage}}
\maketitle
\begin{abstract}
Shear particle acceleration is a promising candidate for the origin of extended high-energy 
emission in extra-galactic jets. In this paper, we explore the applicability of a shear 
model to 24 \xray knots in the large-scale jets of FR II radio galaxies, and study the jet 
properties by modeling the multi-wavelength spectral energy distributions (SEDs) in a 
leptonic framework including synchrotron and inverse Compton - CMB processes.
In order to improve spectral modelling, we analyze \fermi\, data for five sources and
reanalyze archival data of \chandra \ on 15 knots, exploring the radio to X-ray 
connection.
We show that the \xray SEDs of these knots can be satisfactorily modelled by synchrotron 
radiation from a second, shear-accelerated electron population reaching multi-TeV energies.
The inferred flow speeds are compatible with large-scale jets being mildly relativistic.
We explore two different shear flow profiles (i.e., linearly decreasing and power-law) 
and find that the required spine speeds differ only slightly, supporting the notion that
for higher flow speeds the variations in particle spectral indices are less dependent on 
the presumed velocity profile. 
The derived magnetic field strengths are in the range of a few to ten microGauss,
and the required power in non-thermal particles typically well below the Eddington
constraint.
Finally, the inferred parameters are used to constrain the potential of FR II jets 
as possible UHECR accelerators. 
\end{abstract}




\section{Introduction}
Radio galaxies are characterized by the large-scale radio emission on scales from kiloparsec (kpc) to megaparsec (Mpc) energized by the jets launched from their active galactic nuclei (AGNs). 
Based on their observational morphology, they are divided into low-power Fanaroff-Riley type I (FR I) sources and high-power Fanaroff-Riley type II (FR II) sources \citep{Fanaroff1974}.
Kpc-scale jets in radio galaxies have been studied for several decades.
Their multi-wavelength images at radio, optical, and \xray wavelengths commonly consist of bright knots \citep{Kraft02,Clautice16a,Hardcastle2016}.
The radio and optical emission from kpc-scale jets are considered to be produced by synchrotron radiation of electrons, however the origin of the extended \xray emission is still unclear \citep{Harris2006}.

For most knots in FR I jets, the radio, optical, and \xray spectrum can typically be explained by synchrotron radiation from a single population of electrons \citep[e.g.][]{Perlman2001, Hardcastle2001, Sun2018}. 
The detection of the extended TeV emission from the kpc-scale jet of Centaurus A by the High Energy Stereoscopic System (H.E.S.S.) also supports the synchrotron origin of the \xrays emission \citep{HESS20}. 
On the other hand, in FR II jets the \xray emission can exhibit much harder spectra than seen in the radio to optical band, which can not be modeled by the synchrotron radiation from a single population of electrons \citep[e.g.][]{Jester2006,Jester2007}.
It has been proposed that such extended \xray emission could be produced by inverse Compton up-scattering (IC) of 
cosmic microwave background (IC/CMB) photons \citep{Georganopoulos2003, Abdo2010, McKeough2016, Wu2017, Guo2018, 
Zhang2018b}, although a large jet (bulk) Lorentz factor $\Gamma$ would then be required on kpc-scales
\citep[e.g.][]{Tavecchio2000,Celotti2001,Zhang2010, Zhang2018a}.
However, this scenario is challenged by recent polarimetry observations and \gray observations \citep[see also][for a 
review]{Georganopoulos2016}, e.g. in the jets of 3C 273 \citep{Perlman2020, Meyer2014}, PKS 0637-752 \citep{Perlman2020,
Breiding2023}, and PKS 1136-135 \citep{Cara2013,Breiding2023}. 
In an alternative scenario, the hard \xray spectra could be related to the synchrotron radiation of a second electron 
population that is different from the radio-optical emission \citep[e.g.,][]{Jester2006,zhang2009,Georganopoulos2016,
Sun2018}, or to the synchrotron radiation of protons in the extended regions of large-scale jets \citep{Aharonian2002MNRAS, Kundu2014MNRAS}.

A synchrotron origin of \xrays emission requires $\sim100$ TeV electrons, which will cool on a timescale of a few thousand years 
in a typical magnetic field strength of $\sim10~\mu$G. This corresponds to a distance of several hundred pc. 
Thus for jet knots of sizes larger than 1 kpc, a distributed (re)acceleration mechanism is required to maintain 
the diffuse \xray emission in the knots. Shear acceleration is a promising candidate mechanism for this 
\citep{Liu2017,Rieger2019Galax,Wang2021MNRAS,Tavecchio2021}.
In shearing flows, particles can gain energy by elastically scattering off small-scale magnetic field inhomogeneities 
embedded in velocity-shearing layers. The process can in principle be understood as a Fermi-type particle acceleration 
mechanism \citep{Rieger2004a,Rieger2007,Liu2017,Lemoine2019,Rieger2019Galax}. 
The accelerated particle spectra and achievable maximum energies have been extensively studied, and found to be mainly 
depending on the velocity profile and turbulence spectrum \citep[e.g.][]{Liu2017,Webb2018,Webb2019,Webb2020,Rieger2019,Rieger2021,Rieger2022a,Wang2021MNRAS,Wang2023MNRAS}.
Velocity-shearing flows are naturally expected in AGN jets. For example, high-resolution radio imaging and 
polarization studies have indicated the presence of velocity gradients transverse to the main jet axes in FR II 
jets \citep[e.g.,][]{Boccardi2016, Nagai2014}. In general, interaction of a jet with its environment is likely
to excite instabilities and introduce velocity shearing. In fact, our recent 3D relativistic magneto-hydrodynamic 
simulations have shown that shearing layers can be naturally self-generated by a relativistic jet spine interacting 
with its surrounding medium \citep{Wang2023MNRAS}. 

In a previous paper, we have obtained an exact solution for the steady-state particle spectrum within a Fokker-Planck
approach, and used it to successfully reconstruct the observed, diffuse \xray emission in two exemplary sources: the
kpc-scale jet in Cen A (FR I type), and the knots A+B1 and C2 in the jet of 3C 273 (FR II type) \citep{Wang2021MNRAS}.
In this paper, we further explore the application of such a shear acceleration model to a large-sample of \xray knots 
in FR II type jets, and study the jet properties by modeling their multi-wavelength data.
In Section \ref{section:Chandra_data}, we describe the details of the data analysis process and show spectral properties 
for the \xray and \gray spectrum. 
In Section \ref{section:SED and fitting}, we describe the SED modelling in the framework of shear acceleration.
In Section \ref{section:results}, we present the fitting results with this shear acceleration model and discuss 
their implications. The conclusions are given in Section \ref{section:conclusion}.

\section{Data} \label{section:Chandra_data}
We select a sample of eight FR II radio galaxies with clear morphology and wavelength coverage in the data-set of radio-to-X-ray data from 
the \xray jet catalog\footnote{\url{https://hea-www.harvard.edu/XJET}} and the paper \citet{Zhang2018a}, including 
3C 273, 3C 403, 3C 17, Pictor A, 3C 111,  PKS 2152-699, 3C 353, and S5 2007+777.
%
The details of the sources are shown in the following.


\textbf{3C 273:} 3C 273 is an ideal FR II radio galaxy with rich multi-wavelength observations. 
The origin of the hard \xray emission from its knots has been actively debated \citep{Jester2005, Jester2006,Uchiyama2006,Jester2007, Zhang2010, Zhang2018a,wang2020}. 
It is known to host a super-massive black hole (SMBH) of mass of $\sim6.6\times10^{9}\msun$ \citep{Paltani2005}, $\msun$ is the mass of sun. 
The redshift is z = 0.158, such that $1\arcsec$ corresponds to 2.7 kpc \citep{Sambruna2001}. 
The jet is about $20\arcsec$ in \chandra \ observation, which indicates that the projected length 
of the jet can extend over 50 kpc. 
Proper-motion studies provide an upper-limit on the velocity $(\Gamma < 2.9)$ and a viewing angle 
of $\theta \sim$ $7\deg$ \citep{Meyer2016}.
\citet{Marchenko2017} find that the prominent brightness enhancements in the \xray and 
far-ultraviolet jet of 3C 273 can be resolved transversely as extended features with sizes of about 0.5 kpc.
We select five \xray regions, A, B1+B2, B3+C1, C2, and D1+D2H3 to perform the spectral analysis, and 
combine adjacent knots if they are difficult to distinguish.
The radio and optical data are obtained from \citet{Jester2007}, and the \gray data are taken from \citet{Meyer2014}.

\textbf{3C 403:} 
3C 403 is one of the best examples of synchrotron \xray emission from the jet of a powerful narrow-line radio galaxy \citep{Kraft2005}.
The mass of its SMBH is $\sim 1.8 \times 10^{8}\msun$ as estimated from its K-band bulge luminosity
\citep{Vasudevan2010}. 
According to the unified models of FR II radio galaxies, the jets of narrow-line radio galaxies are at a large 
viewing angle of $> 45\deg$ \citep{Kraft2005,Barthel1989}. 
The east jet of 3C 403 includes two significant \xray knots, F1 and F6. 
The measured redshift of the host galaxy is $z = 0.059$, corresponding to a luminosity 
distance of 260.6 Mpc $(1\rm \arcsec = 1.127\ kpc)$. 
The radio and optical data are taken from \citet{Kraft2005} and \citet{Werner2012}. 

\textbf{3C 17:} 
We select two \xray knots (S3.7 and S11.3) in the powerful jet of the radio galaxy 
3C 17. 
The mass of its SMBH is $\sim 5.0 \times 10^{8}\msun$ \citep{Sikora2007}.
The measured redshift of the host galaxy ($z$ = 0.22) corresponds to a conversion scale of 
$1\rm \arcsec$ = 3.47 kpc.
While we assume a synchrotron origin, we note that given the SED shape and unusual character of S11.3, it cannot be excluded that IC/CMB contributes to the emission from this knot \citep{Massaro2009,Rahman2023}.
The radio, optical, and \xray data are taken from \citet{Massaro2009}. 

\textbf{Pictor A:} Pictor A has an unilateral and straight jet in the radio and \xray energy bands \citep{Gentry2015}. 
This source harbors a SMBH of mass $\sim 4.0\times10^{7}\msun$ \citep{Ito2021}.
The measured redshift of the host galaxy is $z = 0.0304$ $(1\rm \arcsec = 0.697\ kpc$ \citep{Hardcastle2016}. 
\citet{Tingay2000} have estimated a viewing angle $\theta \lesssim 51\deg$ based on {\it Very Long Baseline Array} 
(VLBA) observations. 
We select three knots, HST-32, HST-106, and HST-112, with radio, optical, and \xray data taken from \citet{Gentry2015}. 

\textbf{3C 111:} 3C 111 is a typical FR II radio galaxy with a SMBH of mass $\sim 2.0\times 10^{8}\msun$ \citep{Ito2021}. It is 
located at a redshift of $z$ = 0.158, corresponding to a luminosity distance of 215 Mpc.
\chandra \ observations by \citet{Clautice16a} report \xray emission from three knots, K14, K30, and K61 in the northern 
jet. VLBA observations reveal an angle to the line of sight $\theta \lesssim 20\deg$ and a velocity $\sim$ 0.98 for the 
entire jet \citep{Oh2015}.
The radio and optical data are taken from \citet{Clautice16a}, and the \gray data are taken from \citet{Zhang2018a}.

\textbf{PKS 2152-699:} PKS 2152-699 is a well-studied FR II radio galaxy at a redshift of $z$ = 0.0283, corresponding 
to a luminosity distance of 122 Mpc \citep{Ly2005}.
This radio galaxy is one of the brightest sources in the southern sky at 2.7 GHz \citep{Ly2005}. 
\citet{Worrall2012} found a bright knot D about 10$\arcsec$ from the host galaxy and estimate a total time-averaged 
jet power $4\times10^{44}\rm erg\space \ s^{-1}$. 
The radio and optical data for this knot are taken from \citet{Fosbury1998} and \citet{Worrall2012}. 

\textbf{3C 353:} The jet of 3C 353 has three significant \xray knots, E23, E88, and W47 \citep{Kataoka2008}. 
\citet{Swain1998} estimate a rather large viewing angle of $60\deg<\theta<90\deg$ for the whole jet based on {\it Very 
Large Array} (VLA) observations at 8.4 GHz. 
The measured redshift of the host galaxy is $z = 0.0304$, corresponding to a conversion scale of $1\rm \arcsec$ = 
0.60 kpc \citep{Kataoka2008}. 
The radio, optical, and previous \xray data are taken from \citet{Kataoka2008}.

\textbf{S5 2007+777:} The \xray jet of S5 2007+777 exhibits properties of both FR I and FR II radio galaxies \citep{Sambruna2008a}. 
The jet has an angle of $<32\deg$ to the line of sight, and the deprojected jet length significantly exceeds 
150 kpc \citep{Sambruna2008b}. 
The source harbors an SMBH of mass $\sim 2.5\times 10^{7}\msun$ \citep{Wu2002}. 
The measured redshift of its host galaxy is $z = 0.342$, corresponding to a conversion scale of $1\rm \arcsec$ = 
4.80 kpc \citep{Sambruna2008b}. 
We select five \xray knots, including K3.6, K5.2, K8.5, K11.1, and K15.9. 
The radio data and the optical upper limits are taken from \citet{Sambruna2008b}, and the \gray data are taken from \citet{Mondal2019}.

\subsection{\chandra\ data analysis} 
The \chandra\ \xray Observatory launched in 1999, provides high resolution ($< 0.5\rm \arcsec$) \xray imaging and 
spectroscopy in the energy range 0.1 – 10 keV \citep{Weisskopf2002}. 
The Science Instrument Module of \chandra \ has two focal plane instruments, the Advanced CCD Imaging Spectrometer 
(ACIS) and the High Resolution Camera (HRC). 
The ACIS module is used for spectral analysis.
In this paper, the spectral extraction is performed using the CIAO (v4.13) software and the \chandra \ Calibration 
Database (CALDB, v4.9.4). 
The spectral analysis is performed using {\it Sherpa}\footnote{\url{https://cxc.harvard.edu/sherpa/threads/index.html}} tool. 

The \xray data of the jets in our sample are all from the \chandra\ \xray Observatory. 
Owing to the accumulative exposure and the enhanced software tools of \chandra\,, we perform an improved analysis for five sources 3C 111, 3C 403, PKS 2152-699, S5 2007+777, and 3C 273 to derive more accurate spectrometric information.
The observational information of the five FR II radio galaxies is shown in Table~\ref{tab:1} (see the Appendix~\ref{appendix_table}). 
We analyze the \chandra\ ACIS data following the guidance of $Science$ $Threads$\footnote{\url{https://cxc.harvard.edu/ciao/threads/index.html}}. 
In order to reduce the deviations caused by the position offsets of different observations, we perform astrometric corrections. 
The counts image, exposure map, and the weighted PSF map are produced by performing {\it fluximage} and {\it mkpsfmap} tools, respectively.
We obtain the locations of target sources using the {\it wavdetect} tool. 
For observations with more than two times, we perform the cross-matching between the 
reference observation and the others; we use {\it wcs\_match} to produce a transform 
matrix and {\it wcs\_update} tool to update the coordinates of the shorter observation. 
We select the longest exposure observation as a reference.

For the spectral analysis, we perform aperture photometry using {\it specextract} on each knot. 
The locations of the selected regions, the corresponding length and width ($L_{\rm knot}$ and $W_{\rm knot}$) of the knot are listed in Table \ref{tab:regions} (see the Appendix~\ref{appendix_table}), where $L_{\rm knot}$ is the half width at half maximum along the jet, $W_{\rm knot}$ is the half width at half maximum transverse to the jet. 
We use {\it sherpa} package to perform the broadband fitting of multi-observations simultaneously with a single power-law plus the Galactic absorption model.
The flux of knots in our sample are extracted in the $0.3 - 7.0$ keV energy band. 
We keep the absorption column density \nh\, free, and we do not find evidence for significant deviation for all knots if \nh\, is kept frozen. 
The \xray flux densities and reduced chi-square $\chi^{2}$ are listed in Table~\ref{tab:X-ray data}. 
The signals of \xray radiation from some knots are too weak, leading $\chi^{2}$ to deviate from 1, 
and \nh\, hard to be constrained tightly.
The errors of flux and photon index are calculated at 90\% confidence level. 
When the spectral indices for some knots are not convergent due to too few photons, we set them to 
be 1.0. 
The spectral indices of the \xray emission are typically in the range of $0.5 - 1.2$.
The spectral index measurement for knot D1+D2H3 in 3C 273 is $1.20^{+0.27}_{-0.11}$. We note that
\citet{Jester2006} obtained a somewhat harder \xray spectrum between $1.02\pm{0.05}$ and $1.04\pm{0.04}$. 
The difference might partly be related to systematic effects from the corrections 
for ACIS contamination \footnote{\url{https://cxc.harvard.edu/ciao/why/acisqecontamN0015.html}}. 
Since \nh\, is kept free in our analysis, the error on the spectral index is larger.
In general, the spectral indices do not reveal softening along the jet, which indicates that
these FR II jets need an efficient distributed acceleration mechanism to explain the harder 
\xray spectrum. 
Assuming a synchrotron origin of the FR II jet emission, the harder \xray spectra and the 
differences between the spectral indices of $\alpha_{\rm RO}$ and $\alpha_{\rm OX}$ shown 
in Table~\ref{tab:X-ray data} indeed suggest that the radiation from a single electron 
population cannot explain the radio to \xray SED. 


\begin{table*}
\caption{The (re-analyzed) \xray flux of the knots for different energy bands.}
\begin{threeparttable}

\begin{tabular}{|ccccccccccccc}
\hline\hline
Source & Knot &$\nu F_{\nu}$\tnote{a}&$\nu F_{\nu}\tnote{b}$&$\nu F_{\nu}$\tnote{c}&  \nh [$\rm \times 10^{22} /cm^{2}$]\tnote{d}  & $\alpha_{\rm X} $ \tnote{e} & reduced $\chi^{2}$ & $\alpha_{\rm RO}$ & $\alpha_{\rm OX}$ &\\


\hline
3C 273&A&113$\pm$3.73 & 155$\pm$4.67 & 165$\pm$8.67 & < 0.03 &$0.84^{+0.17}_{-0.01}$ & 0.55 & 0.89$\pm$0.01 & 0.76$\pm$0.01\\
 &B1+B2&84.1$\pm$5.74& 102$\pm$6.02 & 109$\pm$9.68 & $0.05\pm{0.03}$ &$0.92\pm{0.09}$&0.74& 0.95$\pm$0.01 & 0.81$\pm$0.02\\
 &B3+C1&23.7$\pm$2.91& 25.2$\pm$2.96 & 23.3$\pm$5.20 & < 0.09 & $1.02^{+0.37}_{-0.17}$ & 0.42 & 0.92$\pm$0.01 & 1.05$\pm$0.03\\
 &C2&24.9$\pm$3.15 & 29.4$\pm$3.54 & 30.1$\pm$3.81 & < 0.02 & $0.94^{+0.15}_{-0.11}$ & 0.49 & 0.97$\pm$0.01 & 1.00$\pm$0.02\\
 &D1+D2H3&54.4$\pm$9.33 & 49.1$\pm$7.74 & 39.7$\pm$8.86 & < 0.11&$1.20^{+0.27}_{-0.11}$& 0.57 & 1.00$\pm$0.01 & 1.13$\pm$0.01\\
3C 403&F1& 1.27$\pm$0.43 & 4.52$\pm$0.76 & 5.83$\pm$2.86 & < 0.86 & 1.00 & 0.99 & 0.58$\pm$0.01 & $1.16^{+0.04}_{-0.05}$ \\
 &F6 & 5.80$\pm$0.94 & 6.07$\pm$0.70 & 6.37$\pm$1.79 & < 0.52 &$0.96^{+1.46}_{-0.71}$& 1.08& 0.71$\pm$0.01 & 1.17$\pm$0.02 \\
3C 111 &K14&3.27$\pm$0.91 & 9.51$\pm$2.69 & 32.6$\pm$14.4 & $0.58^{+0.56}_{-0.41}$ & $0.47^{+0.40}_{-0.38}$& 0.45 & 0.74$\pm$0.01 & $0.96^{+0.04}_{-0.06}$ \\
 &K30& 5.20$\pm$2.02&6.50$\pm$2.65 &9.63$\pm$4.33 & < $0.95$ &$0.72^{+0.79}_{-0.53}$& 0.33 & 0.73$\pm$0.01 & $0.89^{+0.05}_{-0.07}$ \\
 &K61&14.4$\pm$2.31 & 15.0$\pm$1.98 & 16.0$\pm$3.19 & $0.77^{+0.51}_{-0.37}$ & $0.95^{+0.48}_{-0.44}$& 0.53 & $1.00^{+0.02}_{-0.03}$ & $0.81^{+0.06}_{-0.05}$ \\
PKS 2152-699&D&6.74$\pm$1.67 & 5.37$\pm$0.99 &5.26$\pm$2.53& < 0.29 &$1.22^{+0.89}_{-0.42}$& 0.29 & 1.23$\pm$0.01 & 1.09$\pm$0.03\\
S5 2007+777&K3.6&0.56$\pm$0.38 &1.44$\pm$0.79 & < 18.0 & < 1.13 &1.00 & 0.50 & 0.84$\pm$0.02 & -\\
 &K5.2&0.71$\pm$0.45 & 1.90$\pm$0.95 & < 15.8 &-&1.00 & 0.99 &1.10$\pm$0.02& -\\
 &K8.5&2.68$\pm$0.61 & 5.19$\pm$0.91 & 9.76$\pm$3.76 & < 0.56 &$0.94^{+1.14}_{-0.62}$ & 0.45 & 0.72$\pm$0.01 & -\\
 &K11.1&0.61$\pm$0.34 & 1.35$\pm$0.74 & < 10.5 & < 24.4 &1.00 & 0.40 & 0.73$\pm$0.02 & -\\
 &K15.9&1.68$\pm$0.63 & 1.77$\pm$0.54 & 2.22$\pm$1.51 & < 9.88 &1.00 & 0.30 & 0.95$\pm$0.02& -\\
\hline
\hline
\end{tabular}
\begin{tablenotes}
\footnotesize
\item[a] Flux in energy range of $0.3 - 0.8$ keV. 
\item[b] Flux in energy range of $0.8 - 2.5$ keV. 
\item[c] Flux in energy range of $2.5 - 7.0$ keV. 

\item[d] Hydrogen-absorbing column density.
\item[e] For some knots, the photon counts are not high enough to constrain the index, in which case they are set to be 1.\\
Spectral index $\alpha$ and flux are expressed as $F_\nu \propto \nu^{-\alpha}$, $\nu$ denotes the frequency.  
The errors of the flux, \nh, and $\alpha_{\rm X}$ are calculated at a 90\% confidence level. 
The errors of $\alpha_{\rm RO}$ and $\alpha_{\rm OX}$ are calculated based on the error bars in their corresponding references.
\end{tablenotes}
\end{threeparttable}
 
\label{tab:X-ray data}

\end{table*}

\begin{table*}
\centering
\caption{The \gray flux of the knots.}
\begin{threeparttable}
\begin{tabular}{ccccccccc}

\hline
\hline

Source& Time interval (MET ) &$\nu F_{\nu}$\tnote{a}&$\nu F_{\nu}\tnote{b}$&$\nu F_{\nu}$\tnote{c}\\
\hline
Pictor A& 239557417 - 668590613 & 10.4$\pm$0.51 & 3.87$\pm$0.07 & 2.87$\pm$0.71 \\
3C 17& 239557417 - 668305234 & 7.00$\pm$0.02 & 1.21$\pm$0.57 & -\\
PKS 2152-699& 239557417 - 668305234 & 7.00$\pm$0.31 & 1.11$\pm$0.67 & -\\
3C 403& 239557417 - 668631991 & < 0.70 & - & -\\
3C 353& 239557417 - 668933761 & < 0.50 & - & -\\
\hline
\hline
\end{tabular}
\begin{tablenotes}
\item[a] Flux in the energy range of $0.5 - 1.3$ GeV.
\item[b] Flux in the energy range of $1.3 - 3.7$ GeV. 
\item[c] Flux in the energy range of $3.7 - 10$ GeV.\\
MET denotes the Mission Elapsed Time. 
\\
The upper limits are computed at $99\%$ confidence level ($3\sigma$).  
\end{tablenotes}
\end{threeparttable}
\label{tab:Fermi}
\end{table*}




\subsection{ \fermi\ data analysis}
The Fermi Large Area Telescope (\fermi), launched in 2008, is a wide field-of-view imaging \gray telescope covering the energy range from about 20 MeV to more than 300 GeV\footnote{\url{https://fermi.gsfc.nasa.gov/ssc/data/analysis/documentation/Cicerone/Cicerone_Introduction/LAT_overview.html}} \citep{Atwood2009}.
We select \fermi\ Pass 8 data around the sources 3C 17, 3C 353, 3C 403, Pictor A, and PKS 2152-699 regions. The observation time is listed in Table~\ref{tab:Fermi}. 
We use a 10\deg$\times$10\deg\ square region centered at the position of target sources as the region of interest. 
We process the data through the current Fermitools from conda distribution\footnote{\url{https://github.com/fermi-lat/Fermitools-conda/}} together with the latest version of the instrument response function {\it P8R3\_SOURCE\_V3}.
We select the "source" event class in an energy range from $0.5 - 10$ GeV for individual source analysis. 
Both the front and back converted photons are included. 
To exclude time periods when some spacecraft event affected the data quality, we use the recommended expression {\it \rm (DATA\_QUAL > 0) \&\& (LAT\_CONFIG == 1)}. 
To reduce the background contamination from the earth’s albedo, only the events with zenith angles less than 90\deg\ are included. 
We apply the Python module that implements a maximum likelihood optimization technique for a standard binned analysis \footnote{\url{https://fermi.gsfc.nasa.gov/ssc/data/analysis/scitools/python_tutorial.html}}.
In the background model, we include the sources in the \fermi\ ten-year catalog \citep[4FGL-DR2,][]{Ballet2020}.
We use the script make4FGLxml.py\footnote{\url{https://fermi.gsfc.nasa.gov/ssc/data/analysis/user/}} to generate the source model files, and the parameters for the target sources within 9.0{\it \deg}of the center are set free. 
For the diffuse background components, we use the latest Galactic diffuse emission model {\it gll\_iem\_v07.fits} and isotropic extragalactic emission model {\it iso\_P8R3\_SOURCE\_V3\_v1.txt}\footnote{\url{https://fermi.gsfc.nasa.gov/ssc/data/access/lat/BackgroundModels.html}} with their normalization parameters free. 
We assume the target source is a point-like source and has a power-law spectrum. 
To derive the SED, we divided the energy interval into three equal bins in logarithmic space and performed the maximum likelihood fitting in each energy bin. 
For 3C 353 and 3C 403, where the signals are detected with a significance of less than $2\sigma$, we calculated 
the upper limits within $3\sigma$ confidence level. 
The derived flux calculated within $3\sigma$ confidence level is listed in Table~\ref{tab:Fermi}, and the SEDs are shown in Figure~\ref{fig:2}.

\section{SED modeling and fitting} \label{section:SED and fitting} 

In our shear acceleration model, the radio-to-\xray data is explained by synchrotron radiation from two 
populations of electrons \citep{Wang2021MNRAS}.
The low-energy electron population is responsible for the radio to optical emission, and might be 
related to first-order shock or second-order Fermi acceleration processes \citep[e.g.,][]{Rieger2007,
Liu2017,Tavecchio2021}. The high-energy electron population is responsible for the UV-to-\xray observation, 
and thought to be related to shear acceleration.

For simplicity we assume the low-energy population to have an exponential-cutoff power-law shape at $E\geq 
E_{\rm min1}$,  
\begin{equation}
    N_{1}(E)=A_{\rm 1} \left(\frac{E}{E_{0}}\right)^{-\alpha_{\rm 1}}\exp\left[-\left(\frac{E}{E_{\rm cutoff1}}\right)^2\right], 
    \label{eq:injected electron spectrum}
\end{equation} where $A_1$ is the normalization constant, $\alpha_{\rm 1}$ denotes the spectral index 
of the low-energy electrons, $E_{\rm cutoff1}$ is the cut-off energy and $E_{0}$ is set equal to 1 TeV, $E_{\rm min1}$ is the minimum energy of the low-energy electron.
We note that the corresponding exponential shape might be obtained if diffusion of these electrons proceeds in 
the Bohm regime \citep[e.g.,][]{Zirakashvili2007}.

For the shear-accelerated high-energy population, we adopt the exact solution of the steady-state 
Fokker-Planck-type equation at $E\geq E_{\rm min2}$ \citep{Wang2021MNRAS},
\begin{equation}
\label{solq}
N_{2}(E)= A_+E^{-\alpha_+} F_+(E, q)
+ A_-E^{-\alpha_-} F_-(E, q)\,.
\end{equation}
The power-law spectral indices are given by 
\begin{equation}
\alpha_\pm={1-q\over2}\mp\sqrt{{(5-q)^2\over4}+w},\label{eq:spm}
\end{equation}
where the $\alpha_{-}$ component dominates the particle spectrum. $w$ is a dimensionless measure 
of the shear viscosity, while $q$ denotes the power-law index of the turbulent spectrum. Here we adopt 
a Kolmogorov-type turbulence spectrum ($q = 5/3$), which is in general consistent with numerical 
simulations \citep{Wang2023MNRAS}. 
The functions $F_\pm(E, q)$ are defined as
\begin{equation}
F_\pm(E, q)={_1F_1}\left[ {2-\alpha_\pm \over q-1},{2\alpha_\pm\over 1-q};-{6-q\over q-1}
\left(\frac{E}{E_{\rm e,max}}\right)^{q-1}\right], 
\end{equation} 
where $E_{\rm e,max}$ is the cut-off energy, and $_1F_1$ denotes the Kummer's confluent hyper-geometric 
function \citep{Abramowitz1972}. 
The integration constants $A_{\rm \pm}$ can be obtained by the condition $N_{2}(E)\to0$ at $E\to\infty$ 
and the normalization of the spectrum. 

In general, particle acceleration in shearing flows depends on the underlying flow velocity profile. 
Here we explore two different velocity profiles: For a linearly decreasing profile with $\beta(r) = 
\beta_{\rm 0, l}[1-(r/R_{\rm jet})]$, the shear viscosity $w$ can be expressed as \citep{Wang2023MNRAS},
\begin{equation}
w_l= \frac{10\beta_{\rm 0, l}^{2}}{[\zeta\ln(1-\beta_{\rm 0, l}^{2})+2\beta_{\rm 0, l} 
\rm tanh^{-1}(\beta_{\rm 0, l})]^{2}}.  
\label{eq:wlinear}
\end{equation}
where $\beta_{\rm 0, l}$ is the velocity on the jet axis, and $\zeta = W_{\rm sh}/R_{\rm jet}$, is $\sim$ 1/2 for FR II jets following the simulation result, here $R_{\rm jet}$ = $W_{\rm knot}$ denotes the jet radius, and $W_{\rm sh}$ is the width of the shearing region. 

For a power-law type velocity profile with $\beta(r) = \beta_{\rm 0, p}/[1+(5r/R_{\rm jet})^{2}]$, where $\beta_{\rm 0, p}$ denotes the spine velocity, $w$ is of the form \citep{Rieger2022a},
\begin{equation}
w_p= \frac{(6-q) t_{\rm acc}}{t_{\rm esc}} = 
116\langle \beta \rangle \ln^{-2}{(1+\Delta\beta)\over(1-\Delta\beta)},
\label{eq:wpowerlaw}
\end{equation}
where $\Delta\beta = (\beta_{\rm 0, p} - \beta(R_{\rm jet}))/(1-\beta_{\rm 0, p}\beta(R_{\rm jet}))$ is 
the relativistic relative velocity, $t_{\rm acc}$ and $t_{\rm esc}$ are the accelerating and the 
escaping time, respectively, and $\langle \beta \rangle \equiv \textstyle \int_{0}^{R_{\rm jet}} 
\beta(r)dr/\textstyle \int_{0}^{R_{\rm jet}} dr $ < 1 is a weighted, spatial average of the considered 
velocity profile. 
As $\beta_{\rm 0, l}$ and $\beta_{\rm 0, p}$ approach the speed of light ($\beta \rightarrow 1$), 
one obtains $w$ {\rm →} 0, and the spectral index becomes $\alpha_{\rm -}=3 - q=\frac{4}{3}$, which 
implies that in a jet with ultra-relativistic velocity, the spectral index of the accelerated 
electrons $\alpha_{-}$ becomes independent of the shape of the velocity profile \citep{Webb2018,
Webb2019,Rieger2019,Rieger2022a}.

To ensure that electrons can be effectively accelerated, two requirements need to be satisfied: 
(1) The scattering time is smaller than the acceleration time; 
(2) the acceleration time is smaller than the cooling time. 
Combining these two requirements, the corresponding cut-off energy of electrons and the resultant 
maximum energy of synchrotron photons can be expressed as \citep{Wang2021MNRAS},
\begin{align}
&E_{\rm e,{max}}=0.7B_{1}^{-2} W_{\rm sh,0.1}^{-1}w^{-1/2}(1+f)^{-1}~{\rm PeV},\label{eq:true_gmax}\\
&E_{\rm \gamma, max}=82.3B_{1}^{-3} W_{\rm sh,0.1}^{-2}w^{-1}(1+f)^{-2} ~{\rm keV}, \label{Emaxsyn}
\end{align}
where $B_{1}$ = $B/10~\mu$G is the magnetic field, $f=U_{\rm rad}/U_{\rm B}$ denotes the energy 
density ratio between the radiation field and the magnetic field with $U_{\rm rad}=4.13\times
10^{-13}(1+z)^4 $erg cm$^{-3}$ for the CMB and $ U_{\rm B}=B^{2}/8\pi$, and $W_{\rm sh, 0.1} 
= W_{\rm sh}/0.1 {\rm kpc}$ is the width of the shearing region, respectively. We note that
Eq.~(\ref{eq:true_gmax}) is formally related to the mean acceleration timescale and thus provides
an conservative lower limit to the acceleration efficiency. In ultra-relativistic flows, 
significantly higher energies might be achieved.

We also take into account IC scattering with CMB photons by the two populations of electrons, as well
as the absorption caused by the extragalactic background light (EBL) following the model \citet{Dom2011} 
to fit the \fermi~\gray data.
We note, however, that \fermi~can not resolve the \gray emission region of FR II radio galaxies, and hence, the \gray emission may originate from the jet or the core. Therefore, the \gray data is only treated as upper limits for the knots in the modeling.

\begin{table*}

\centering
\caption{Derived parameters from our SED fits}
\begin{threeparttable}
\begin{tabular}{cccccccccc}
\hline
\hline
Source&Knot&$W_{\rm e,1}$&$W_{\rm e,2}$&$\alpha_{\rm 1}$&$w$&$E_{\rm cutoff1}$&$B$&$E_{\rm min2}$&$E_{\rm min1}$\\ 
 & &$[\times 10^{56} \rm erg]$&$[\times 10^{54} \rm erg]$& & &$[\rm TeV]$&$[\mu \rm G]$&$[\rm TeV]$&$[\rm GeV]$\\ \hline \hline
{\it 3C} 273 & A &$48.9^{+0.35}_{-0.69}$&$167^{+1.71}_{-1.37}$&$2.8^{+0.03}_{-0.02}$&$4.45^{+0.10}_{-0.11}$&$1.69^{+0.15}_{-0.18}$&$3.22^{+0.09}_{-0.14}$&$0.47^{+0.08}_{-0.07}$&8.0\\
 &B1+B2&$541^{+7.31}_{-5.90}$&$77.5^{+22.4}_{-23.2}$&$2.5\pm{0.01}$&$5.19^{+0.06}_{-0.07}$&$1.09\pm{0.01}$&$3.37\pm{0.02}$&$1.50^{+0.03}_{-0.04}$&8.0\\
 &B3+C1&$81.8^{+53.2}_{-35.0}$&$32.6^{+6.88}_{-3.26}$&$2.6\pm{0.01}$&$5.70^{+0.33}_{-0.32}$&$1.74\pm{0.04}$&$2.99^{+0.08}_{-0.07}$&$2.16^{+0.31}_{-0.24}$&8.0\\
 &C2&$59.2^{+0.51}_{-0.70}$&$16.3^{+0.42}_{-0.60}$&$2.5\pm{0.01}$&$5.00^{+0.19}_{-0.21}$&$0.94\pm{0.02}$&$4.80^{+0.35}_{-0.45}$&$0.80^{+0.46}_{-0.18}$&8.0\\
 &D1+D2H3&$425^{+12.4}_{-17.3}$&$54.2^{+2.01}_{-3.02}$&$2.7\pm{0.01}$&$5.45\pm{0.07}$&$1.29\pm{0.01}$&$5.03\pm{0.10}$&$0.75\pm{0.03}$&8.0\\
 \hline
 3C 403&F1&$4.89\pm{0.13}$&$8.62^{+15.2}_{-6.30}\times10^{-3}$&$1.8\pm{0.01}$&$4.02^{+7.84}_{-3.32}$&$0.46\pm{0.02}$&$3.24^{+0.38}_{-0.33}$&$53.4^{+19.2}_{-42.1}$&2.5\\
 &F6&$2.57^{+0.57}_{-0.41}$&$0.05\pm{0.01}$&$2.0\pm{0.01}$&$2.53^{+1.37}_{-0.83}$&$0.75^{+0.05}_{-0.04}$&$5.11^{+0.59}_{-0.61}$&$29.2^{+12.2}_{-9.73}$&2.5\\
3C 17&$\rm S3.7$&$0.16^{+0.07}_{-0.04}$&$1.65^{+1.14}_{-0.88}\times10^{-3}$&$2.6\pm{0.01}$&$3.48^{+0.35}_{-0.32}$&$1.64^{+0.13}_{-0.12}$&$13.4^{+2.13}_{-1.83}$&$14.8^{+2.26}_{-2.63}$&2.0\\
 &S11.3&$1.41^{+1.12}_{-0.59}$&$2.63^{+4.66}_{-2.62}\times10^{-3}$&$2.8^{+0.05}_{-0.07}$&$2.19^{+0.83}_{-0.79}$&$1.94^{+0.53}_{-0.43}$&$3.98^{+1.57}_{-1.21}$&$13.5^{+11.1}_{-9.83}$&2.0\\
Pictor A&HST-32&$0.55^{+2.07}_{-1.35}$&$0.08^{+0.05}_{-0.03}$&$2.4\pm{0.01}$&$2.77^{+0.26}_{-0.24}$&$10.6^{+1.67}_{-1.76}$&$4.60^{+0.60}_{-0.61}$&$14.5^{+8.13}_{-8.60}$&4.0\\
 &HST-106&$1.46^{+0.04}_{-0.05}\times10^{-2}$&$0.02\pm{0.01}$&$2.5^{+0.07}_{-0.04}$&$5.34^{+0.85}_{-0.65}$&$5.77^{+1.97}_{-1.67}$&$4.56^{+0.72}_{-0.58}$&$3.54^{+0.92}_{-0.96}$&4.0\\
 &HST-112&$0.05\pm{0.01}$&$2.89^{+0.57}_{-0.86}\times10^{-3}$&$2.0\pm{0.03}$&$3.26^{+2.18}_{-1.27}$&$6.50^{+1.10}_{-0.93}$&$4.36^{+1.10}_{-0.86}$&$27.7^{+23.3}_{-16.7}$&4.0\\
3C 111&K14&$1.10^{+0.27}_{-0.35}$&$0.13^{+0.05}_{-0.13}$&$2.4\pm{0.03}$&$1.79^{+0.76}_{-0.56}$&$2.89^{+0.77}_{-0.62}$&$4.22^{+0.76}_{-0.64}$&$55.0^{+28.1}_{-33.5}$&5.5\\
 &K30&$2.81^{+2.56}_{-1.12}$&$0.52^{+0.10}_{-0.09}$&$2.4^{+0.32}_{-0.01}$&$3.12^{+2.79}_{-1.75}$&$20.2^{+7.86}_{-8.46}$&$3.16^{+0.95}_{-0.71}$&$7.65^{+25.2}_{-6.72}$&5.5\\
 &K61&$1.52^{+0.42}_{-0.33}$&$0.04^{+0.05}_{-0.01}$&$3.0^{+0.27}_{-0.08}$&$5.32^{+2.04}_{-1.41}$&$17.5^{+21.5}_{-10.4}$&$3.89^{+1.00}_{-0.72}$&$24.4^{+16.5}_{-17.4}$&5.5\\
PKS 2152-699& D&$0.50\pm{0.03}$&$1.37^{+0.53}_{-0.25}\times10^{-2}$&$2.4^{+0.04}_{-0.06}$&$6.62^{+4.22}_{-2.31}$&$8.33^{+14.4}_{-2.30}$&$2.24^{+0.50}_{-0.45}$&$35.0^{+34.6}_{-25.5}$&10.0\\ \hline
3C 353&E23&$1.91^{+0.98}_{-0.71}$&$0.04^{+0.01}_{-0.02}$&$2.4\pm{0.25}$&$3.84^{+3.33}_{-1.69}$&$13.2^{+17.7}_{-12.8}$&$1.62^{+1.81}_{-0.83}$&$55.3^{+29.2}_{-40.1}$&2.5\\
 &E88&$3.61^{+1.10}_{-1.13}$&$3.42^{+13.8}_{-3.41}\times10^{-3}$&$2.5^{+0.20}_{-0.17}$&$10.3^{+6.71}_{-6.84}$&$23.9^{+11.4}_{-15.3}$&$1.40^{+0.82}_{-0.45}$&$41.4^{+38.4}_{-29.9}$&2.5\\
 &W47&$18.1^{+70.5}_{-13.8}$&$0.05^{+0.22}_{-0.04}$&$2.4^{+0.21}_{-0.29}$&$2.10^{+1.01}_{-0.65}$&$26.2^{+26.4}_{-17.0}$&$2.34^{+1.50}_{-0.94}$&$47.1^{+36.8}_{-30.7}$&2.5\\
S5 2007+777&K3.6&$1.20^{+0.12}_{-0.14}$&$0.09^{+0.06}_{-0.07}$&$2.7\pm{0.03}$&$3.28^{+1.18}_{-1.21}$&$24.5^{+17.0}_{-17.6}$&$8.70^{+1.80}_{-1.55}$&$45.7^{+34.0}_{-29.5}$&2.8\\
 &K5.2&$1.65^{+0.12}_{-0.11}$&$0.17^{+0.11}_{-0.10}$&$3.2\pm{0.03}$&$3.13^{+1.23}_{-1.10}$&$23.1^{+17.3}_{-16.0}$&$8.00^{+1.71}_{-1.40}$&$22.0^{+49.5}_{-12.4}$&2.8\\
 &K8.5&$8.58^{+0.86}_{-0.47}$&$0.02^{+1.25}_{-0.01}$&$2.4\pm{0.02}$&$3.62^{+4.42}_{-2.30}$&$1.51^{+4.12}_{-0.66}$&$3.24^{+0.75}_{-0.58}$&$26.8^{+37.2}_{-18.7}$&2.8\\
 &K11.1&$5.39^{+1.30}_{-0.76}$&$0.35^{+0.87}_{-0.34}$&$2.4^{+0.03}_{-0.02}$&$2.72^{+3.57}_{-1.77}$&$17.2^{+20.4}_{-12.3}$&$1.04^{+0.19}_{-0.18}$&$22.0^{+27.1}_{-15.2}$&2.8\\
 &K15.9&$8.80^{+5.78}_{-2.86}$&$0.11^{+1.42}_{-0.10}$&$2.9^{+0.03}_{-0.04}$&$3.10^{+1.32}_{-1.84}$&$12.9^{+11.8}_{-8.35}$&$3.84^{+0.76}_{-0.86}$&$24.2^{+16.8}_{-15.6}$&2.8\\
 \hline
 \end{tabular}
\begin{tablenotes}
\footnotesize
\item[ ] The subscript 1 denotes the parameters of the low-energy electrons, and the parameters with subscript 2 denotes the parameters of the high-energy population. 
\end{tablenotes}
\end{threeparttable}

\label{tab:fitting} 

\end{table*}

\begin{table*}

\centering

\caption{Derived parameters of the jet dynamics.}
\begin{threeparttable}
\begin{tabular}{ccccccccccccc}
\hline
\hline
Source&Knot&${\beta_{\rm 0, l}}$ & $\beta_{\rm 0, p}$& $\alpha_{-}$& $E_{\rm e,max}$ & $P_{\rm knot}/L_{\rm edd}$&Doppler factor\\
 & & & & &[\rm TeV]& &\\
\hline \hline
\textit 3C 273&A&$0.88\pm{0.01}$&$0.85\pm{0.01}$&$2.4\pm{0.02}$&$140^{+12.9}_{-7.0}$&$1.26\pm{0.02}\times 10^{-3}$&1\\
 &B1+B2&$0.86\pm{0.01}$& $0.82\pm{0.01}$&$2.5\pm{0.01}$&$123^{+2.09}_{-1.73}$ &$1.40^{+0.03}_{-0.02}\times 10^{-2}$&1\\
 &B3+C1&$0.83\pm{0.01}$&$0.79\pm{0.01}$&$2.5\pm{0.06}$&$178^{+12.9}_{-12.3}$ &$2.50^{+1.69}_{-1.08}\times 10^{-3}$&1\\
 &C2&$0.86\pm{0.01}$&$0.83\pm{0.01}$&$2.5^{+0.03}_{-0.04}$&$92.0^{+11.9}_{-18.7}$&$2.06\pm{0.03}\times 10^{-3}$&1\\
 &D1+D2H3&$0.84\pm{0.01}$&$0.80\pm{0.01}$&$2.5\pm{0.01}$&$74.9^{+3.53}_{-3.23}$&$8.89\pm{0.08}\times 10^{-3}$&1\\
  \hline
3C 403&F1&$0.90^{+0.09}_{-0.25}$& $0.87^{+0.12}_{-0.31}$&$2.3^{+1.72}_{-0.74}$&$154^{+242}_{-89.1}$ &$1.64^{+0.23}_{-0.49}\times 10^{-2}$&1\\
 &F6&$0.97^{+0.02}_{-0.06}$&$0.94^{+0.03}_{-0.06}$&$2.0^{+0.28}_{-0.19}$&$78.8^{+37.6}_{-29.9}$ &$9.27^{+2.32}_{-1.97}\times 10^{-3}$&1\\ 
3C 17& S3.7 &$0.93^{+0.01}_{-0.02}$&$ 0.90\pm{0.02}$& $2.1\pm{0.07}$&$64.5^{+26.1}_{-18.6}$&$2.62^{+1.25}_{-0.67}\times 10^{-4}$&$2.72^{+0.30}_{-0.28}$\\
 &S11.3&$0.98\pm{0.01}$&$0.95\pm{0.03}$&$1.9^{+0.18}_{-0.19}$&$483^{+614}_{-257}$ &$1.53^{+1.26}_{-0.66}\times 10^{-3}$&$5.03^{+8.45}_{-1.85}$\\
Pictor A&HST-32&$0.96\pm{0.01}$&$ 0.93\pm{0.01}$&$2.0^{+0.06}_{-0.05}$&$146^{+53.3}_{-35.7}$ &$1.66^{+0.58}_{-0.46}\times 10^{-2}$&$0.70\pm{0.08}$\\ 
 &HST-106&$0.85^{+0.02}_{-0.04}$&$ 0.81^{+0.03}_{-0.04}$&$2.5^{+0.15}_{-0.12}$&$107^{+40.6}_{-31.8}$ &$4.10^{+1.40}_{-1.90}\times 10^{-4}$&1\\ 
 &HST-112&$0.94^{+0.05}_{-0.10}$&$0.91^{+0.05}_{-0.10}$&$2.1^{+0.27}_{-0.42}$&$255^{+235}_{-127}$ &$1.43^{+0.41}_{-0.37}\times 10^{-3}$&1\\
3C 111&K14&$0.99^{+0.01}_{-0.07}$&$0.97^{+0.02}_{-0.08}$&$1.8^{+0.41}_{-0.33}$&$319^{+523}_{-157}$ &$2.12^{+0.71}_{-0.64}\times 10^{-3}$& $2.02^{+0.90}_{-2.00}$\\
 &K30&$0.94^{+0.05}_{-0.12}$&$0.91^{+0.07}_{-0.13}$&$2.0^{+0.52}_{-0.39}$&$408^{+544}_{-224}$ &$5.18^{+5.31}_{-2.46}\times 10^{-3}$&1\\
 &K61&$0.85^{+0.06}_{-0.08}$&$0.81^{+0.07}_{-0.09}$&$2.5^{+0.34}_{-0.26}$&$123^{+85.6}_{-55.1}$ &$3.09^{+0.11}_{-0.01}\times 10^{-3}$&1\\
PKS 2152-699&D  & $0.80^{+0.09}_{-0.13}$ & $0.75^{+0.11}_{-0.16}$&$2.7^{+0.62}_{-0.40}$&$851^{+643}_{-378}$& - &1 \\
 \hline
3C 353&E23&$0.91^{+0.07}_{-0.13}$&$0.88^{+0.08}_{-0.15}$&$2.2^{+0.58}_{-0.35}$&$2413^{+4367}_{-1892}$& - & 1\\
 &E88&$0.69^{+0.24}_{-0.13}$&$0.61^{+0.29}_{-0.17}$&$3.2^{+0.83}_{-1.12}$&$1427^{+2265}_{-857}$& - & 1\\
 &W47&$0.98^{+0.01}_{-0.04}$&$0.95^{+0.02}_{-0.04}$&$1.9^{+0.22}_{-0.15}$&$1125^{+1747}_{-738}$& - & $0.28^{+0.34}_{-0.12}$\\
S5 2007+777&K3.6&$0.94^{+0.03}_{-0.07}$&$0.91^{+0.05}_{-0.06}$&$2.1^{+0.23}_{-0.26}$&$15.2^{+12.6}_{-6.12}$& $2.31^{+0.35}_{-0.40}\times 10^{-2}$& 1\\
 &K5.2&$0.94^\pm{0.05}$&$0.91^{+0.05}_{-0.06}$&$2.1\pm{0.24}$&$18.2^{+14.2}_{-7.58}$& $4.25^{+0.50}_{-0.54}\times 10^{-2}$& 1\\
 &K8.5&$0.92^{+0.07}_{-0.17}$ &$0.89^{+0.09}_{-0.20}$&$2.2^{+0.76}_{-0.50}$&$109^{+71.6}_{-69.3}$& $0.22^{+0.04}_{-0.05}$& 1 \\
 &K11.1&$0.96^{+0.03}_{-0.15}$&$0.93^{+0.05}_{-0.08}$&$2.1^{+0.32}_{-0.23}$&$290^{+243}_{-116}$& $0.14\pm{0.04}$ & 1\\
 &K15.9&$0.95^{+0.04}_{-0.06}$&$0.92^{+0.07}_{-0.06}$&$2.1^{+0.26}_{-0.41}$&$67.8^{+89.3}_{-25.8}$ & $0.23^{+0.17}_{-0.08}$& 1\\

\hline
\end{tabular}
\begin{tablenotes}
\footnotesize
\item []
{$P_{\rm knot}$ is not been calculated for 3C 353 and PKS 2152-699 as the unknown of mass of SMBH.} 
\end{tablenotes}
\end{threeparttable}
\label{tab:environment}
\end{table*}

The fitting of multi-wavelength SEDs is performed with the open-source code Naima \citep{Zabalza2015}, which allows Markov Chain Monte Carlo (MCMC) fitting using emcee package \citep{Foreman13}. 
To reduce the number of free parameters in our model, we fix $E_{\rm min1}$ (list in Table \ref{tab:fitting}) based on the minimum frequency of the radio data and use 
the same value for different knots in the same jet. 
The total energy of the lower-energy electron population $W_{\rm e,1}$, the total energy of the high-energy electron population $W_{\rm e,2}$, $\alpha_{\rm 1}$, $w$, $E_{\rm cutoff1}$, $B$, and the minimum energy of the high-energy electron population $E_{\rm min2}$ are left as free parameters. 

\section{Results} \label{section:results} 
\begin{figure*}
\flushleft
\includegraphics[scale=0.29]{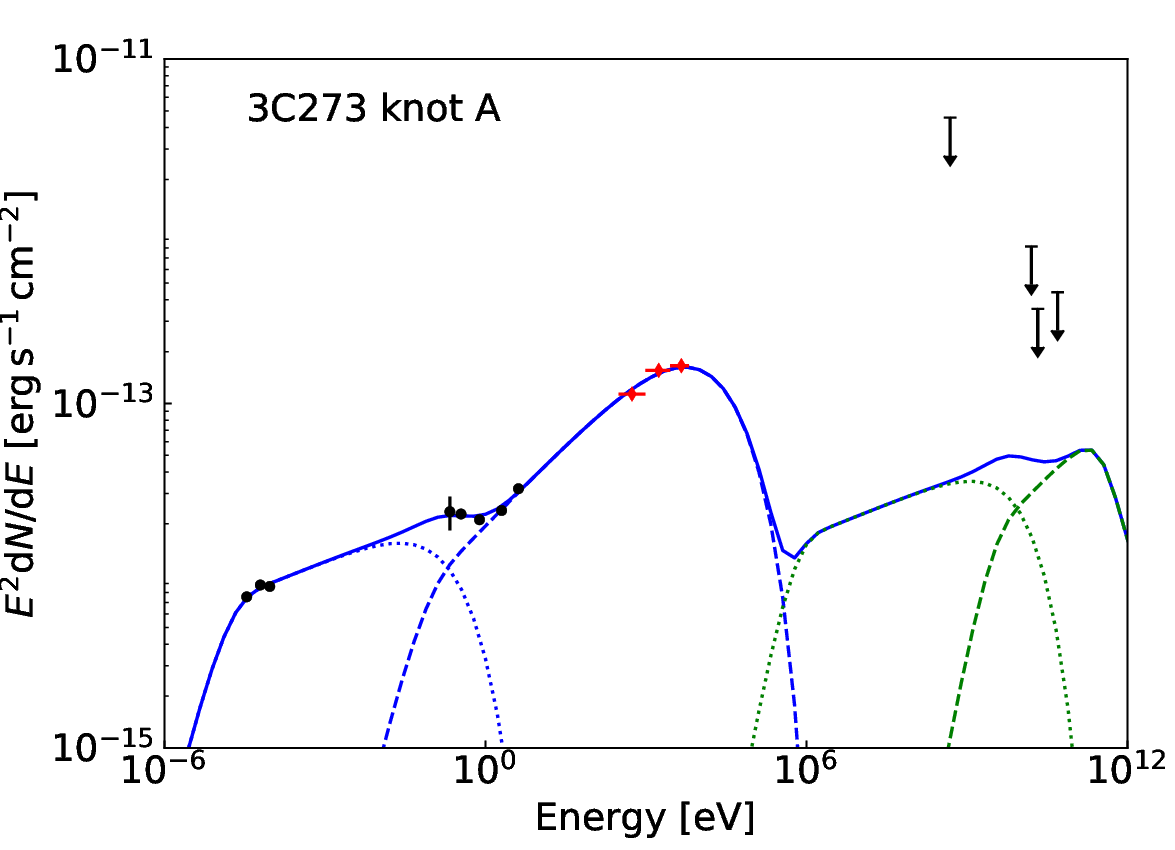}
\includegraphics[scale=0.29]{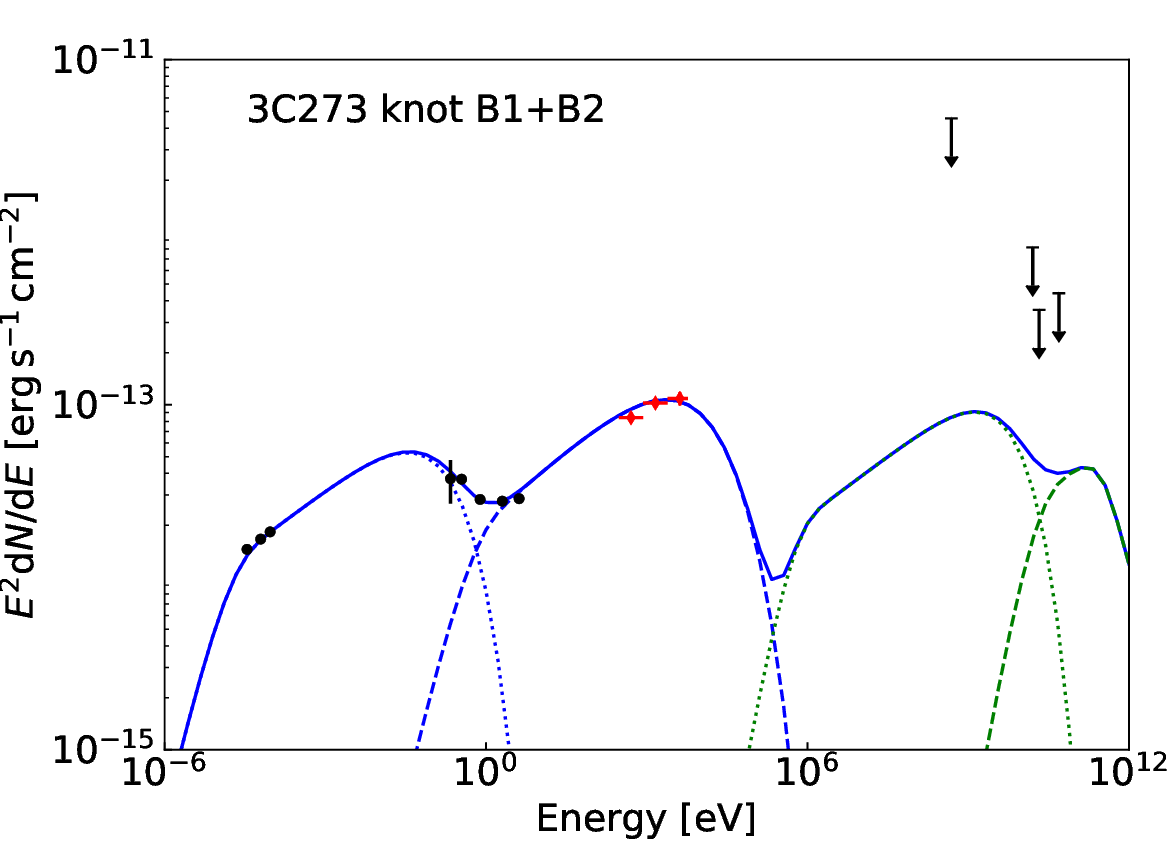}
\includegraphics[scale=0.29]{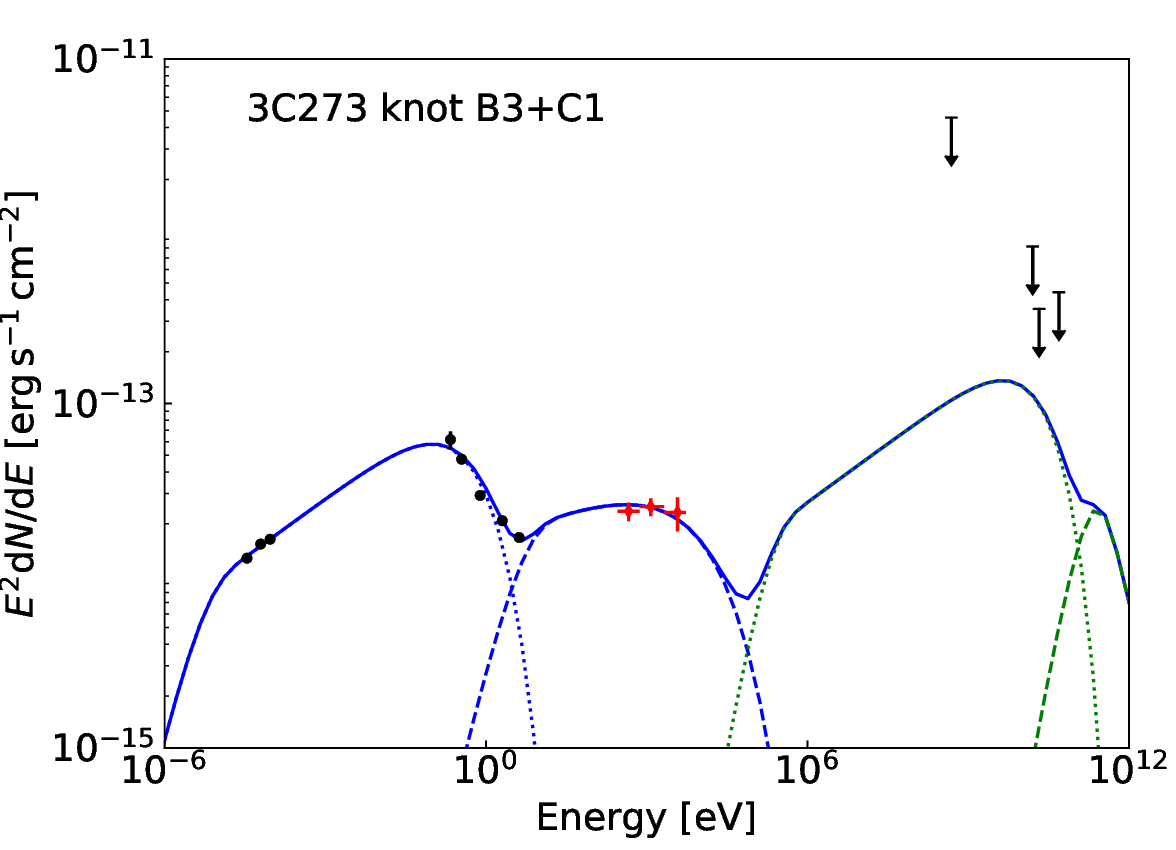}
\includegraphics[scale=0.29]{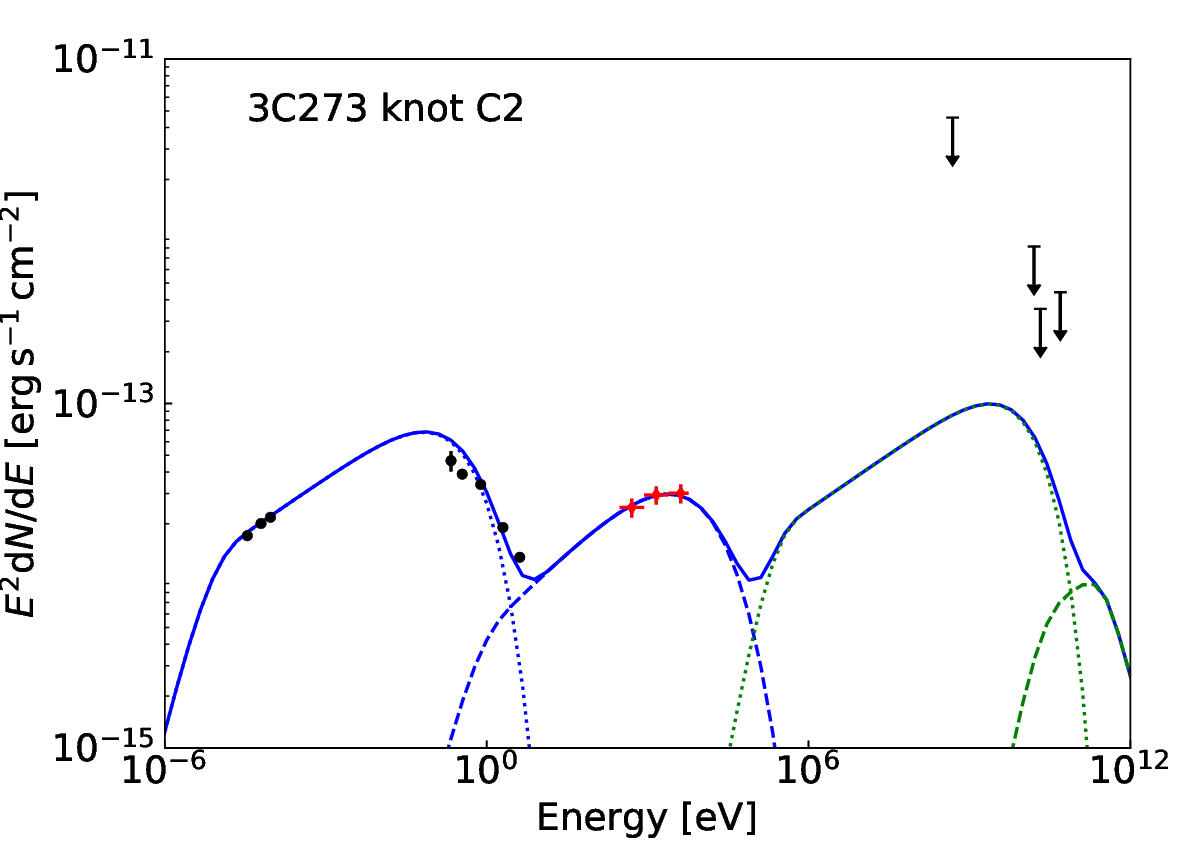}
\includegraphics[scale=0.29]{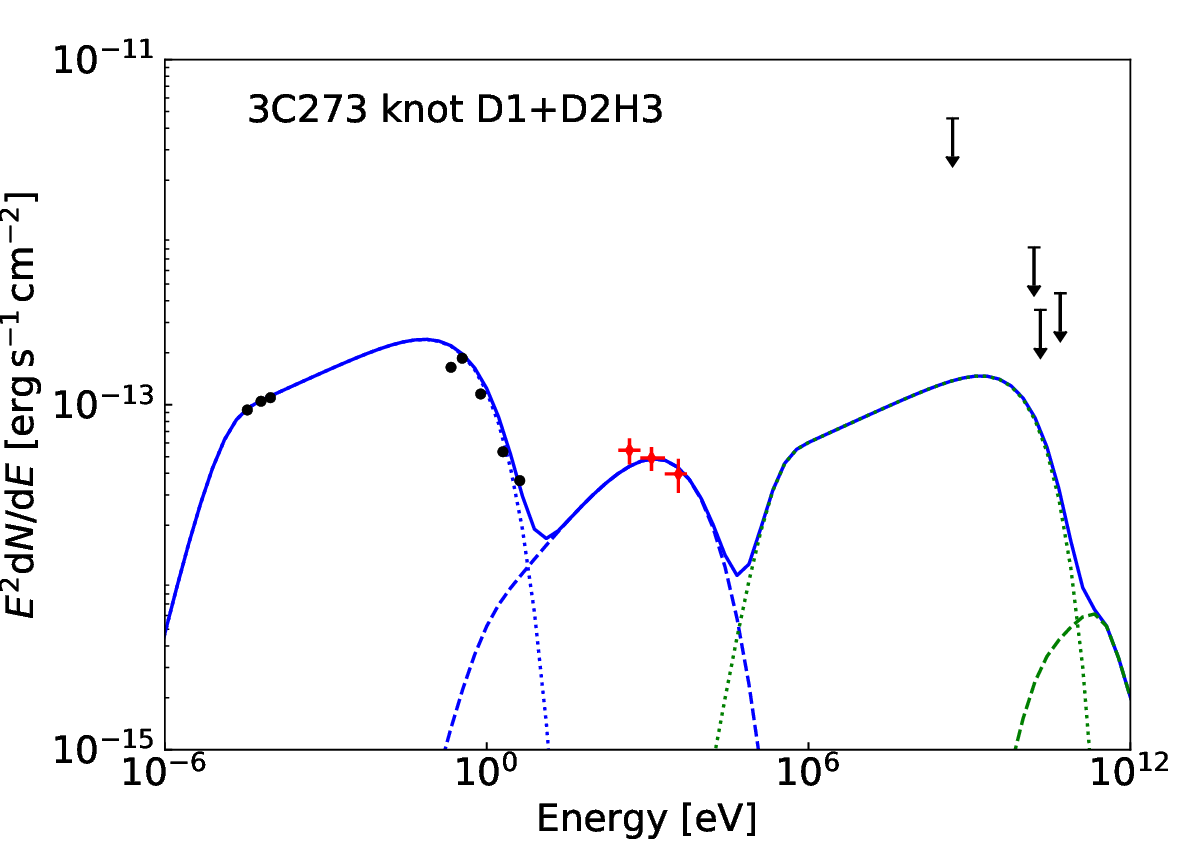}
\caption {SEDs of the \xray knots in 3C 273. 
The solid lines denote the total non-thermal emission from two electron populations. 
The blue and green dotted lines represent the synchrotron and IC/CMB radiation of the 
low-energy electrons, respectively.
The blue and green dashed lines are the synchrotron and IC/CMB radiation of the shear 
acceleration population, respectively.
The red points and upper limits are the data that we analyzed in this paper. 
The black points and upper limits are taken from the references, see details in Section~\ref{section:Chandra_data}. 
}
\label{fig:1}
\end{figure*}

\begin{figure*}
\flushleft
\includegraphics[scale=0.29]{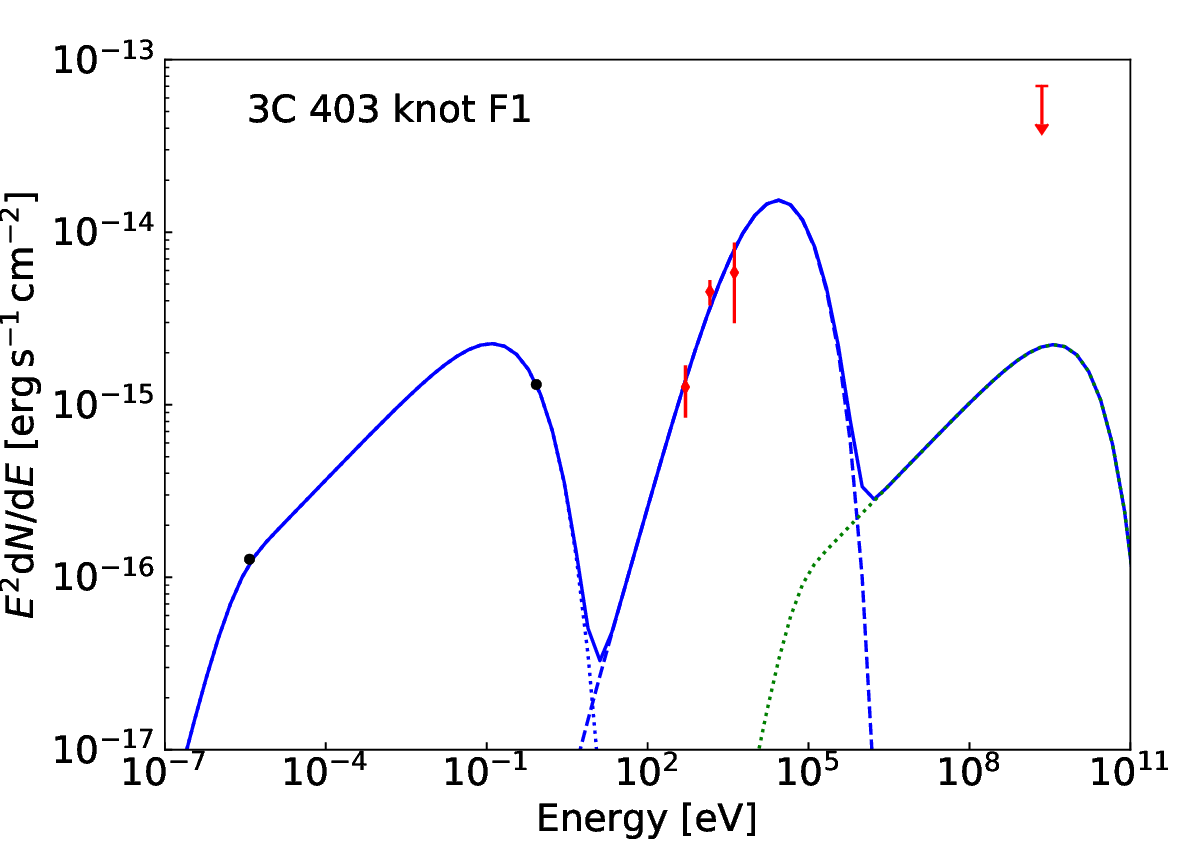}
\includegraphics[scale=0.29]{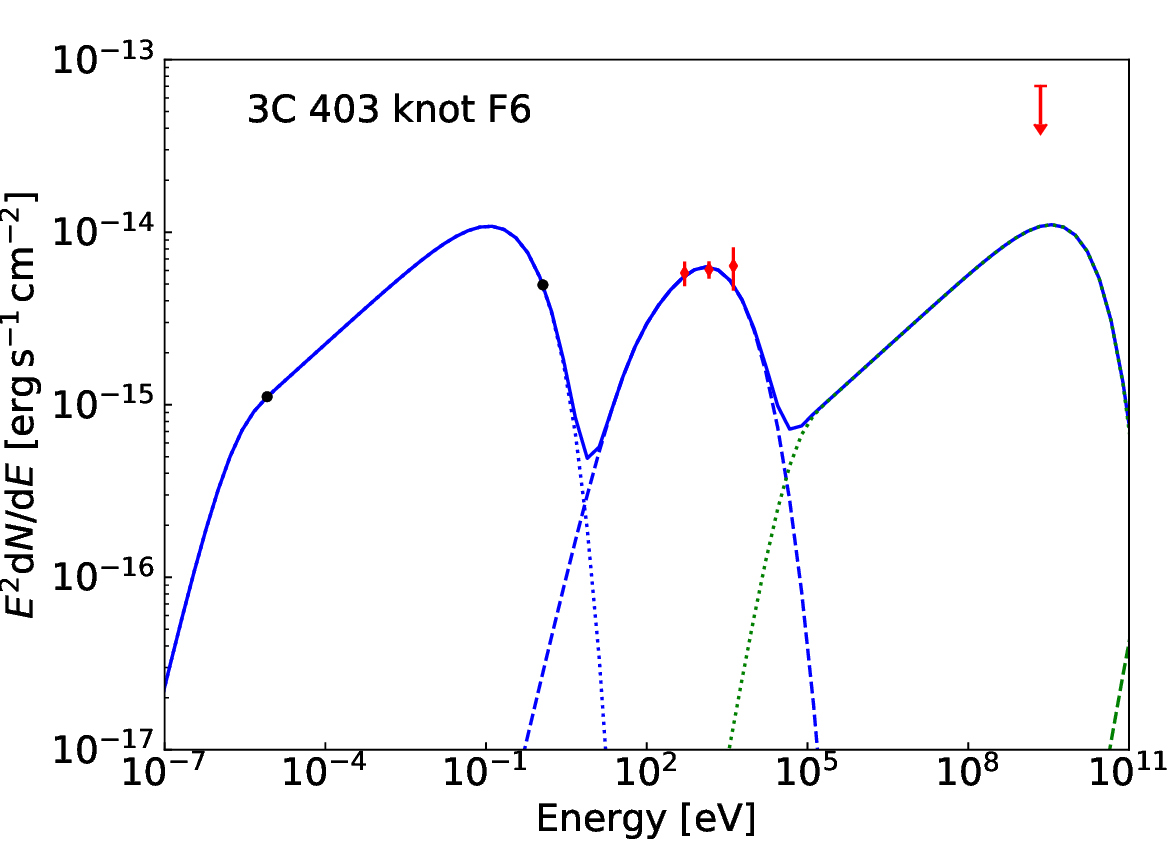}
\includegraphics[scale=0.29]{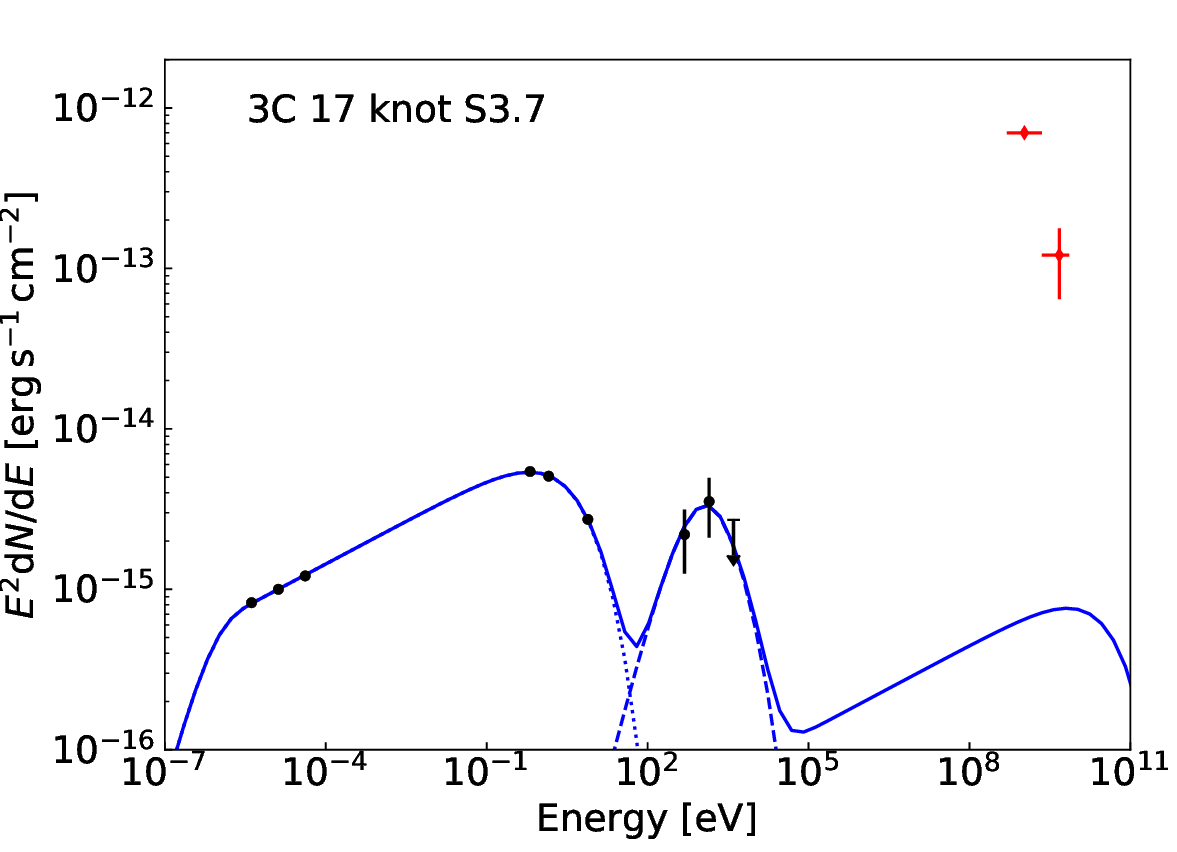}
\includegraphics[scale=0.29]{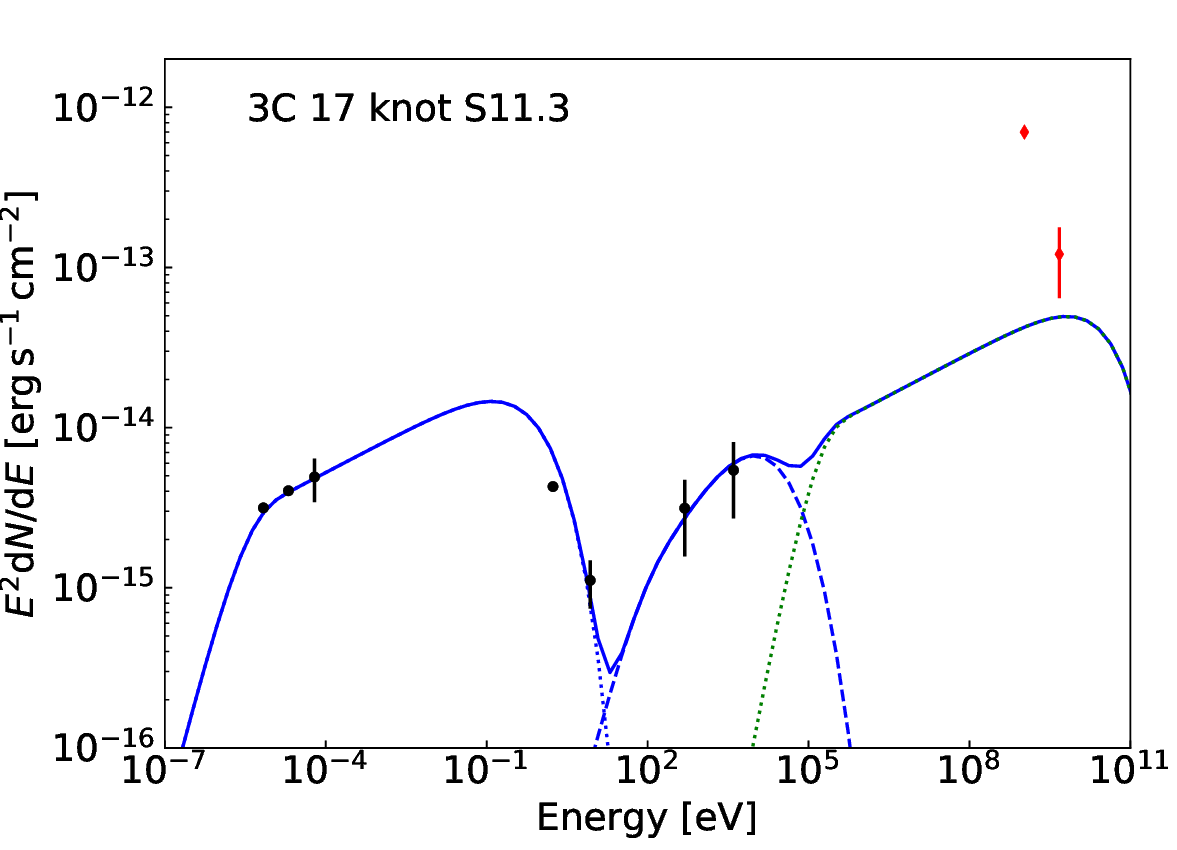}
\includegraphics[scale=0.29]{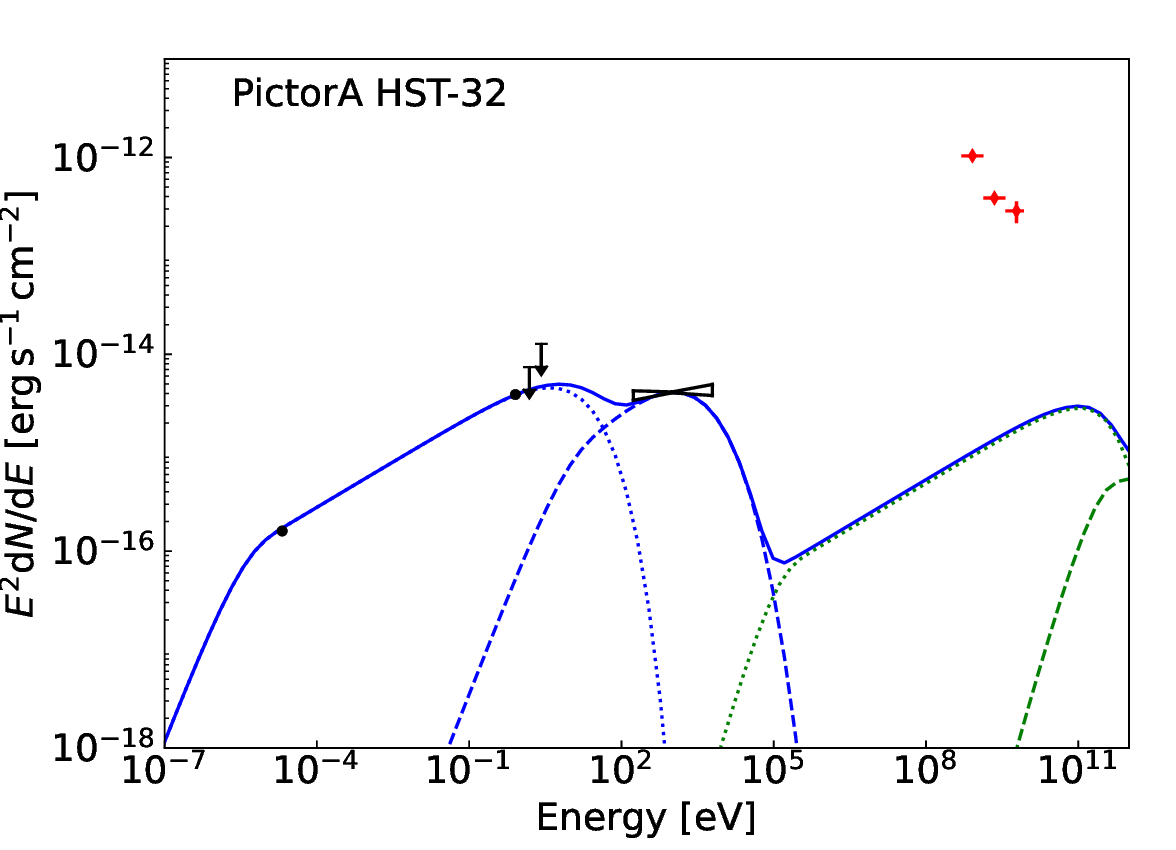}
\includegraphics[scale=0.29]{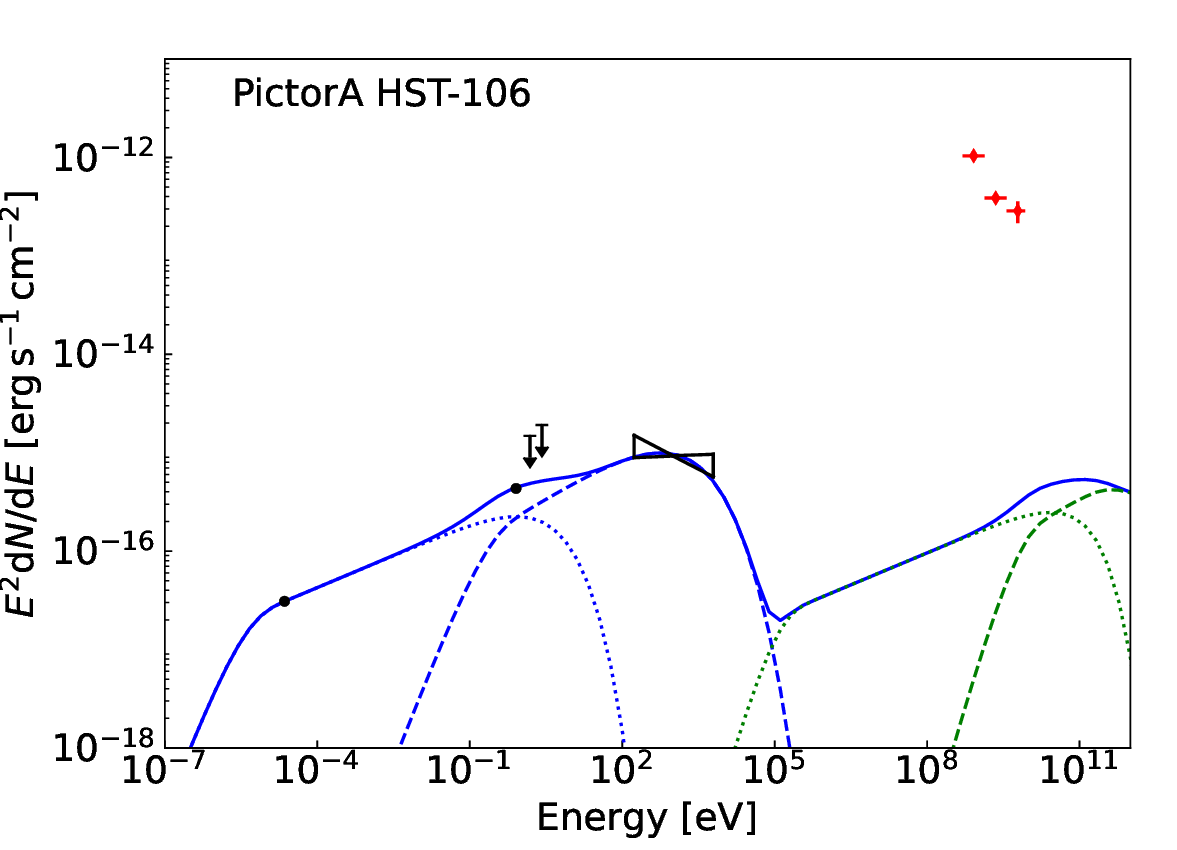}
\includegraphics[scale=0.29]{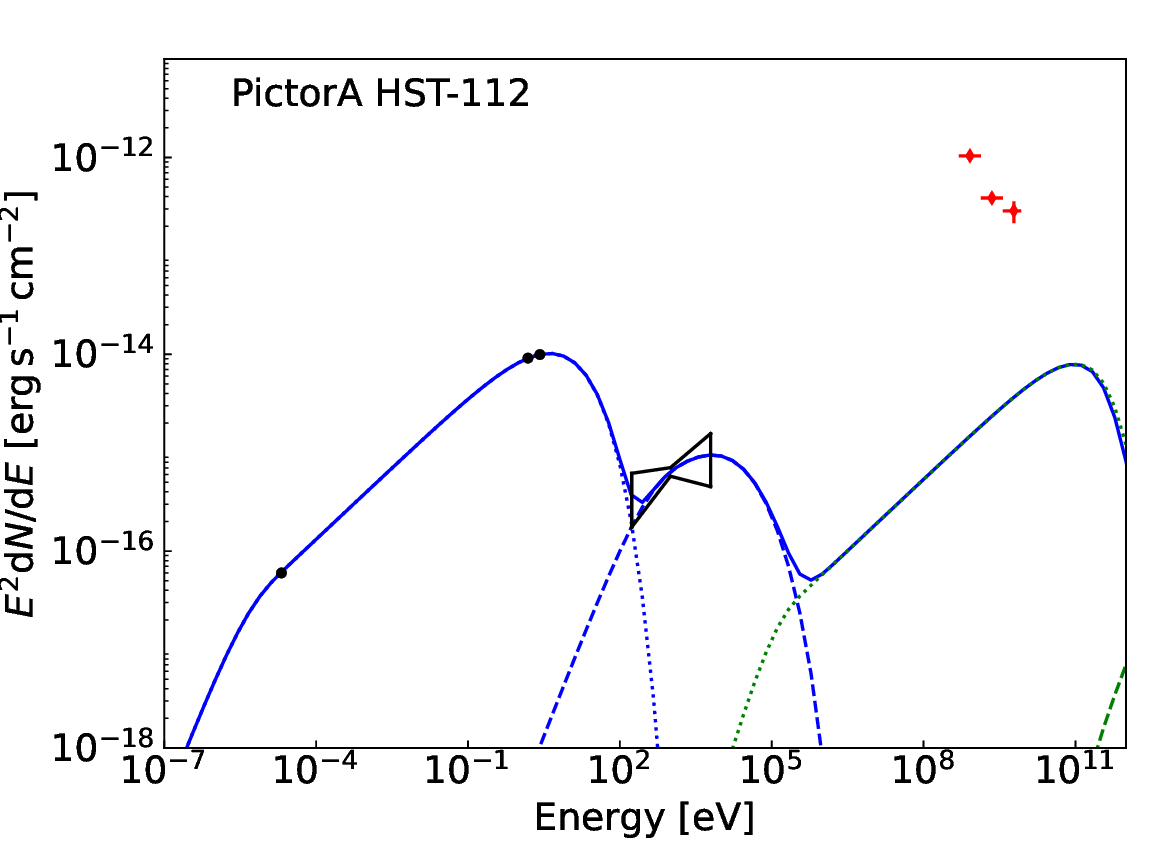}
\includegraphics[scale=0.29]{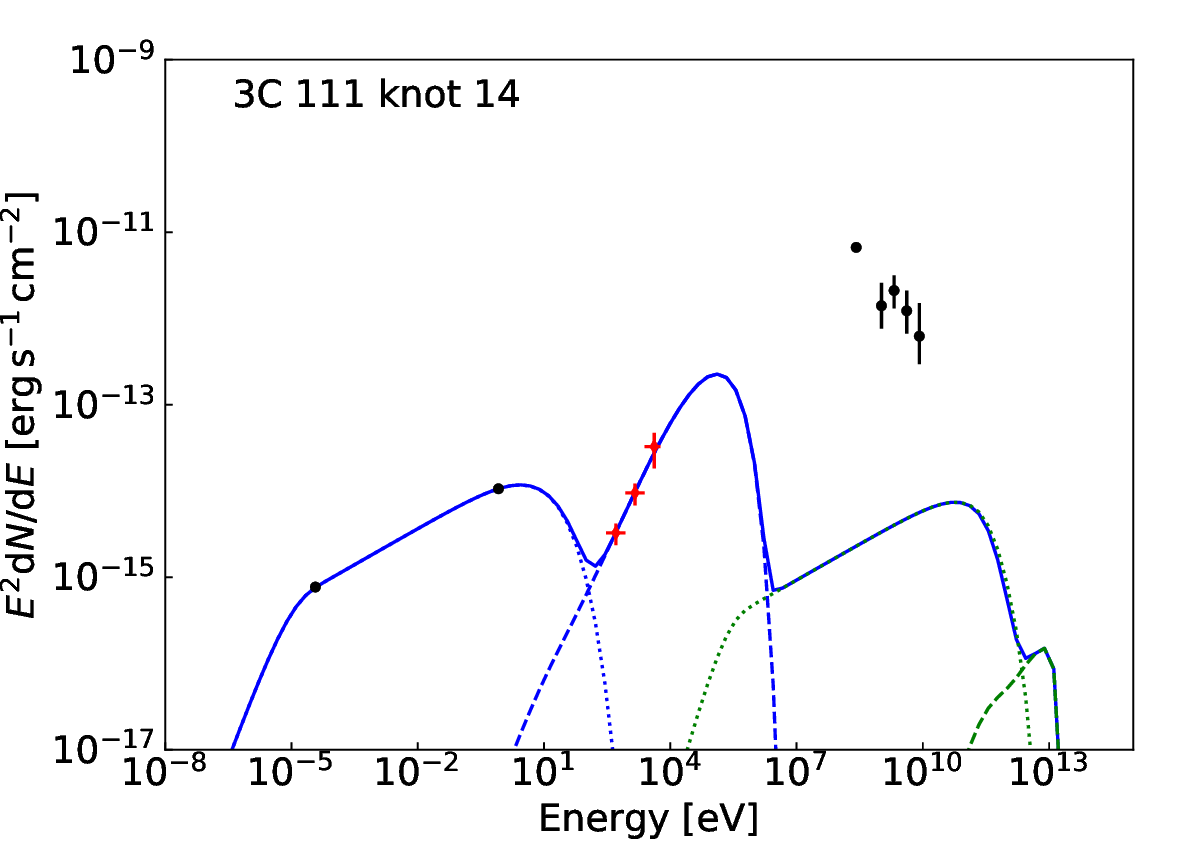}
\includegraphics[scale=0.29]{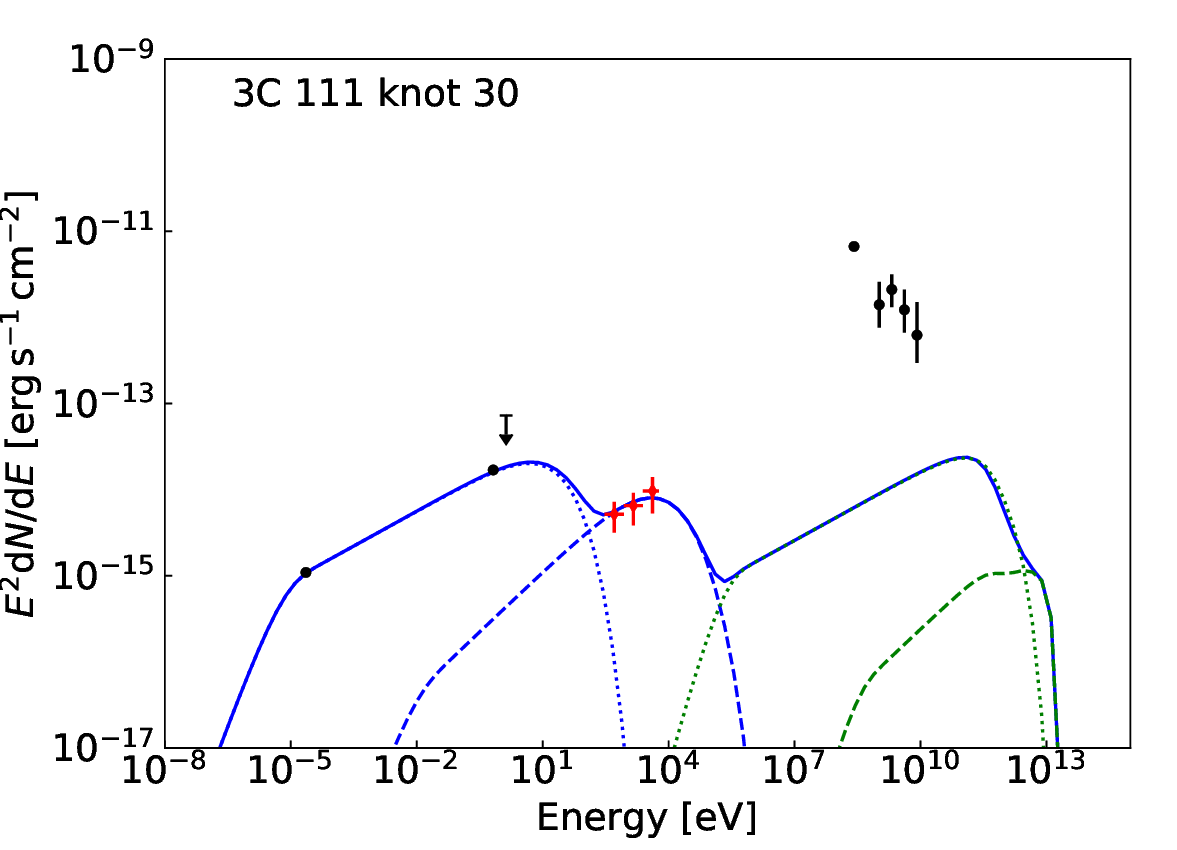}
\includegraphics[scale=0.29]{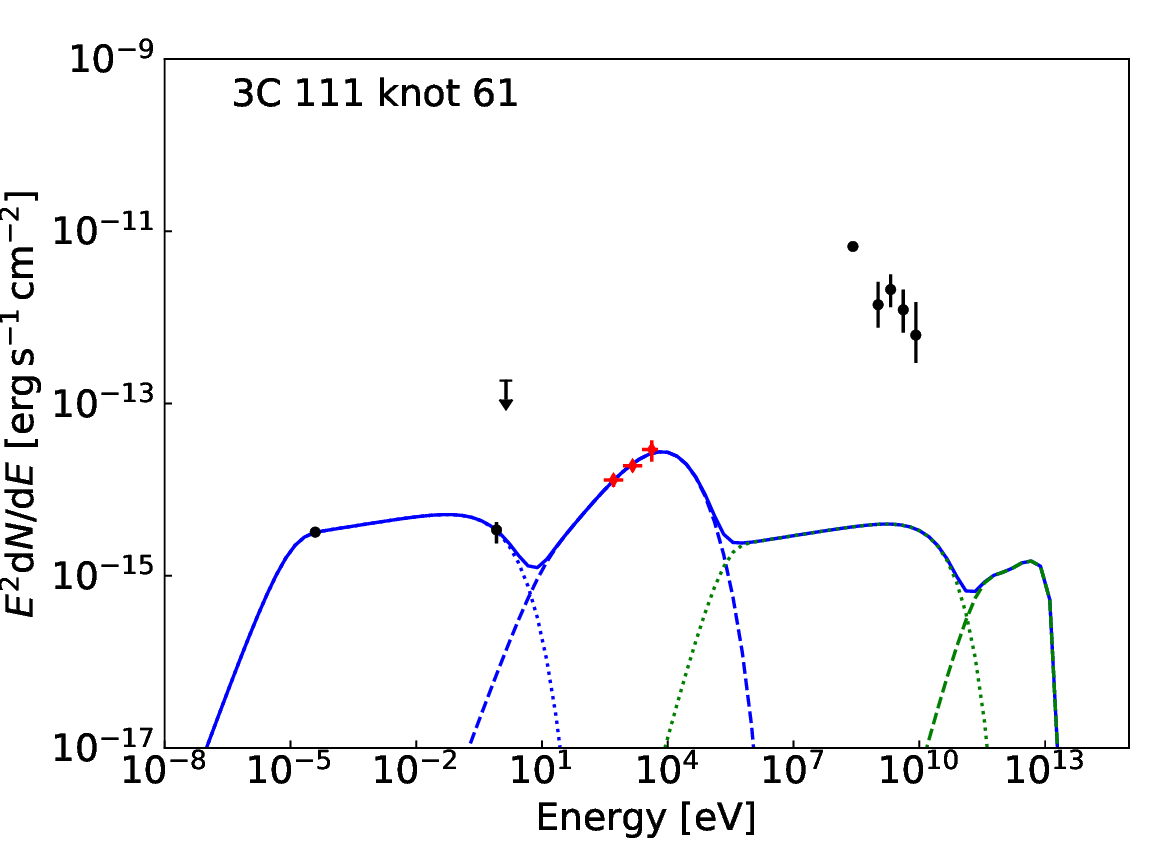}
\includegraphics[scale=0.29]{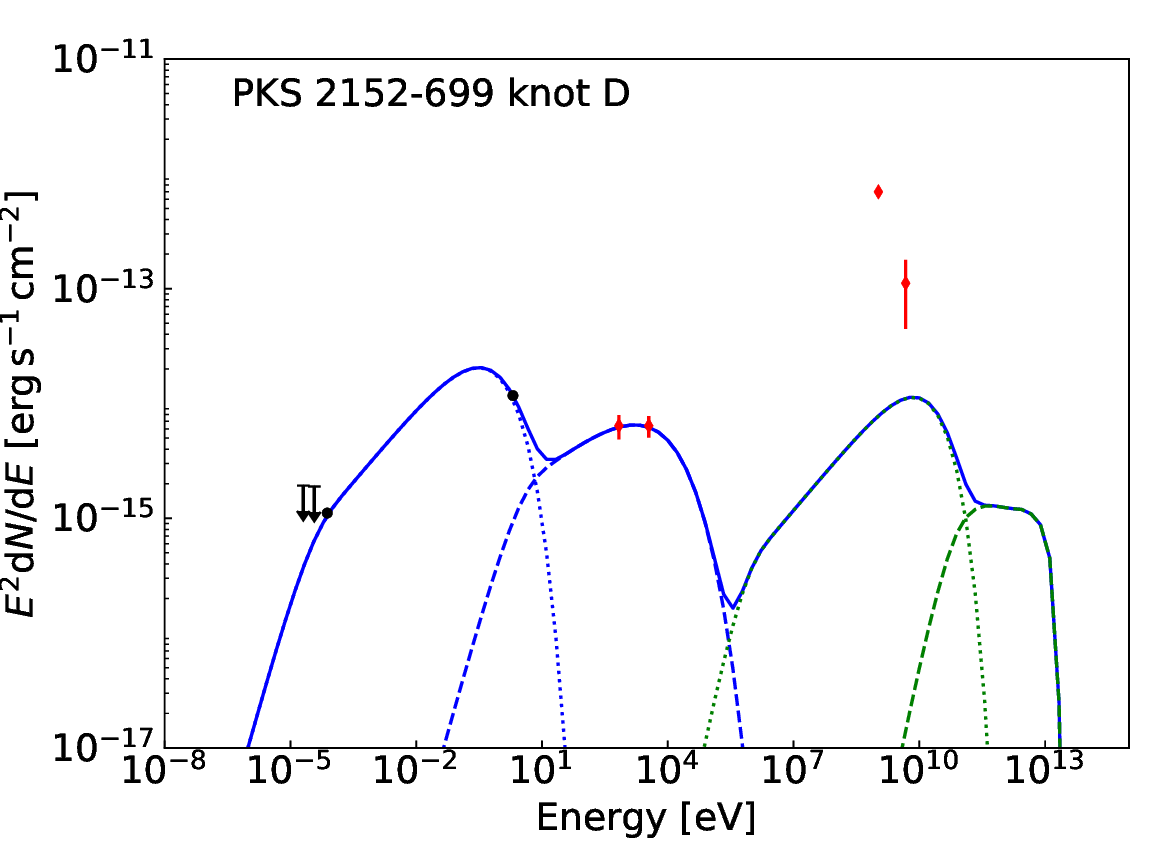}
\caption {Fitting results and measured SEDs of the \xray knots in  3C 403, 3C 17, Pictor A, 3C 111, and PKS 2152-699. 
The styles of data and lines are the same as in Figure~\ref{fig:1}.}
\label{fig:2}
\end{figure*}

\begin{figure*}
\flushleft
\includegraphics[scale=0.29]{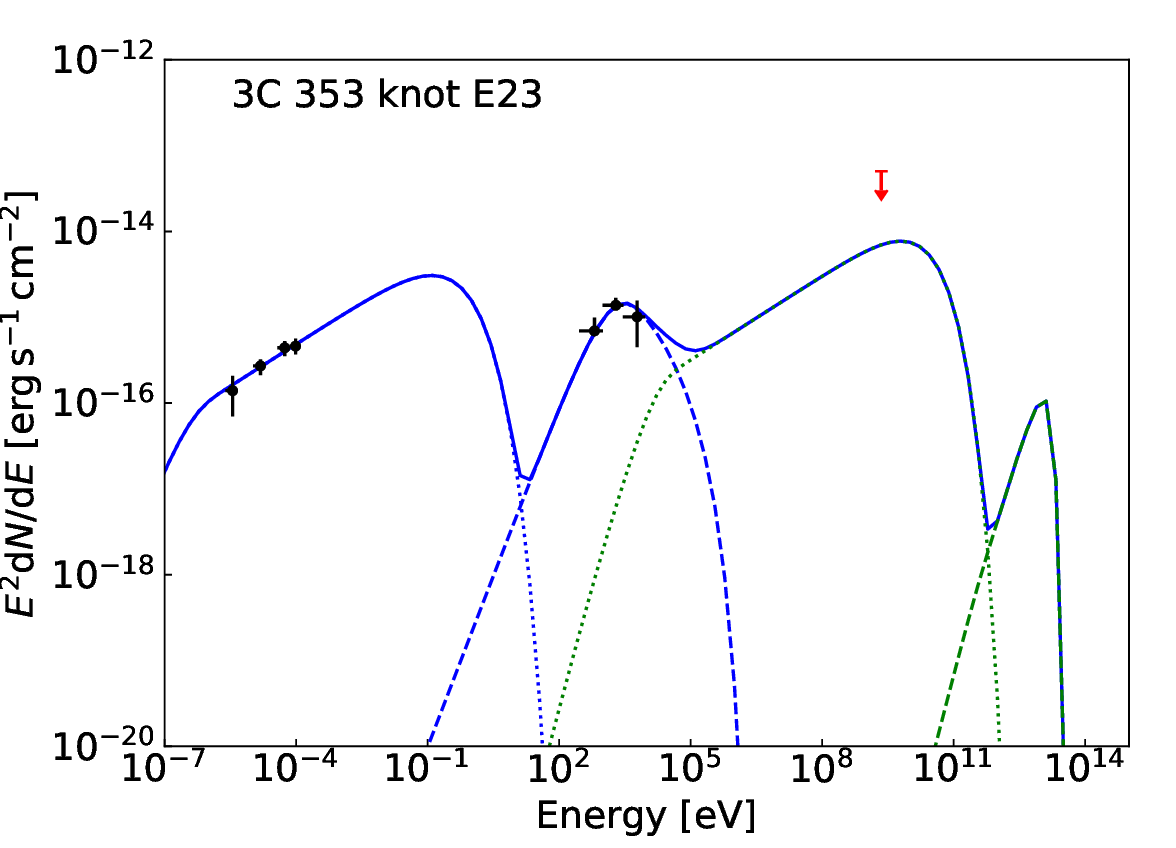}
\includegraphics[scale=0.29]{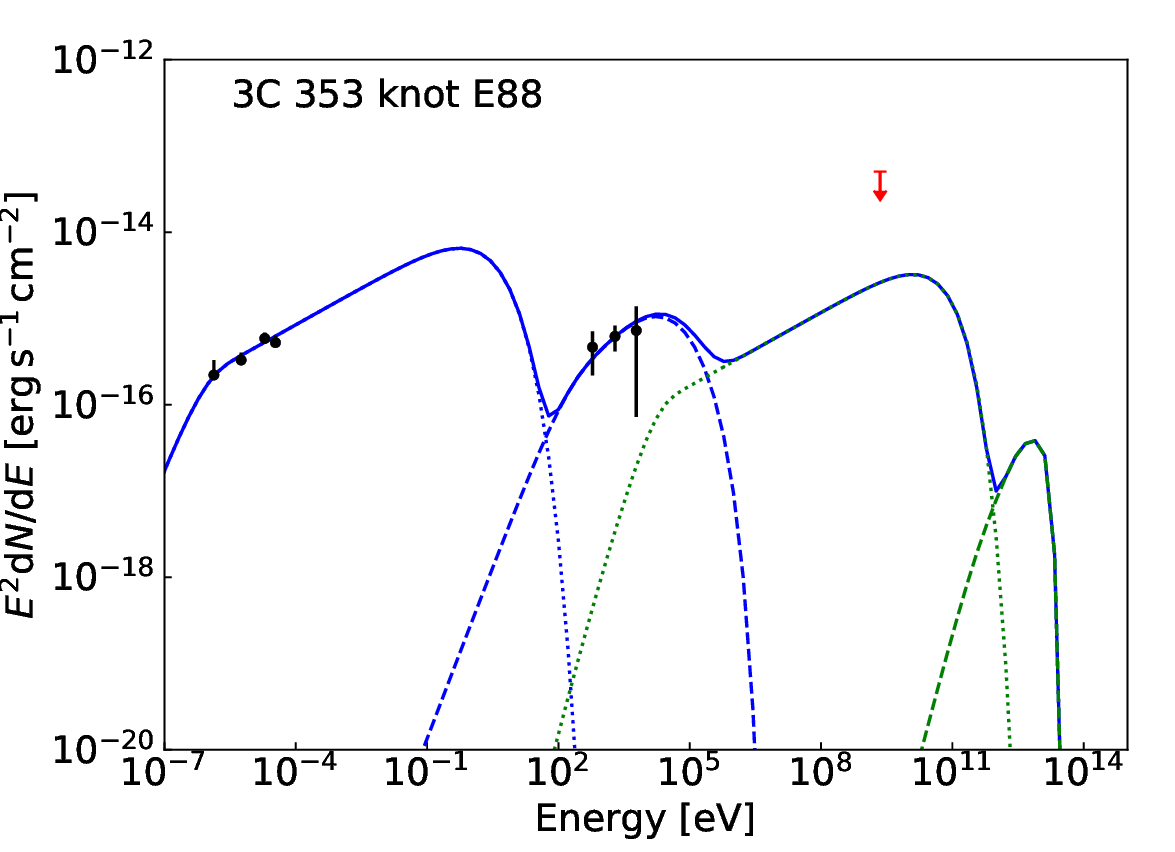}
\includegraphics[scale=0.29]{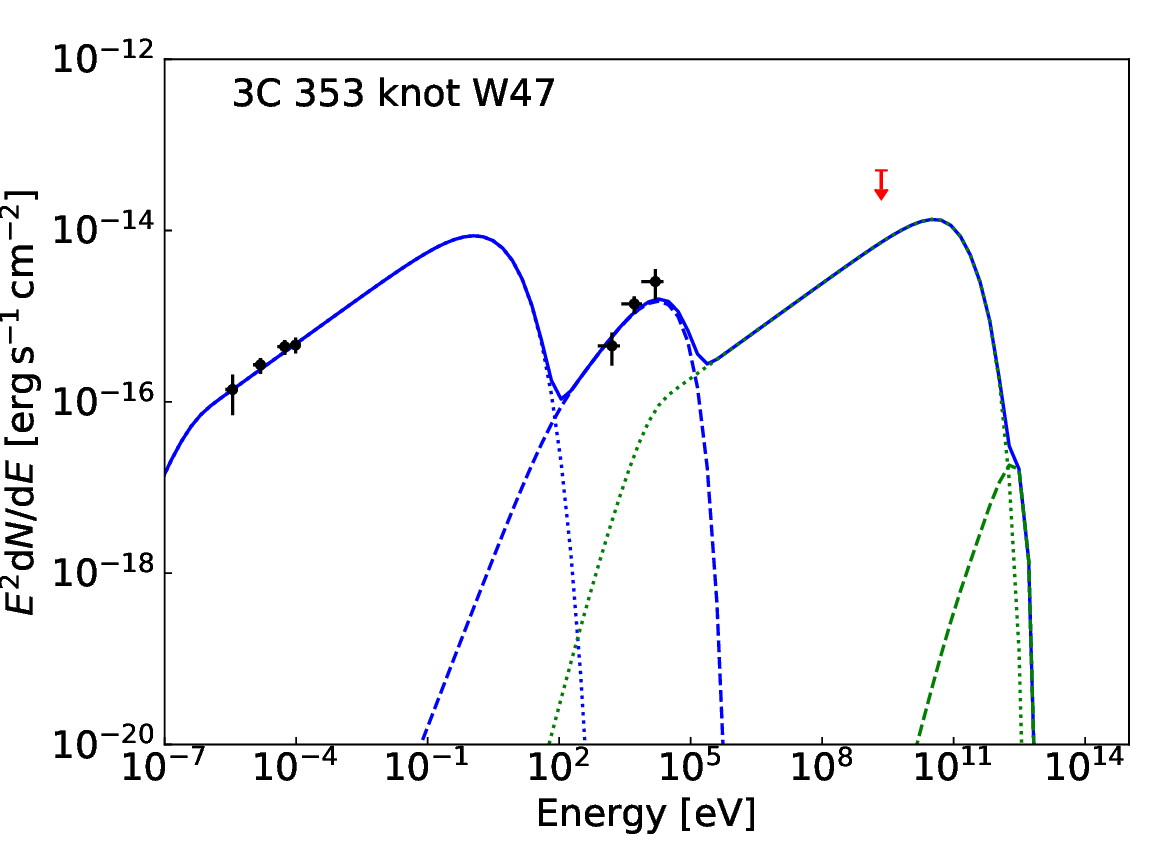}
\includegraphics[scale=0.29]{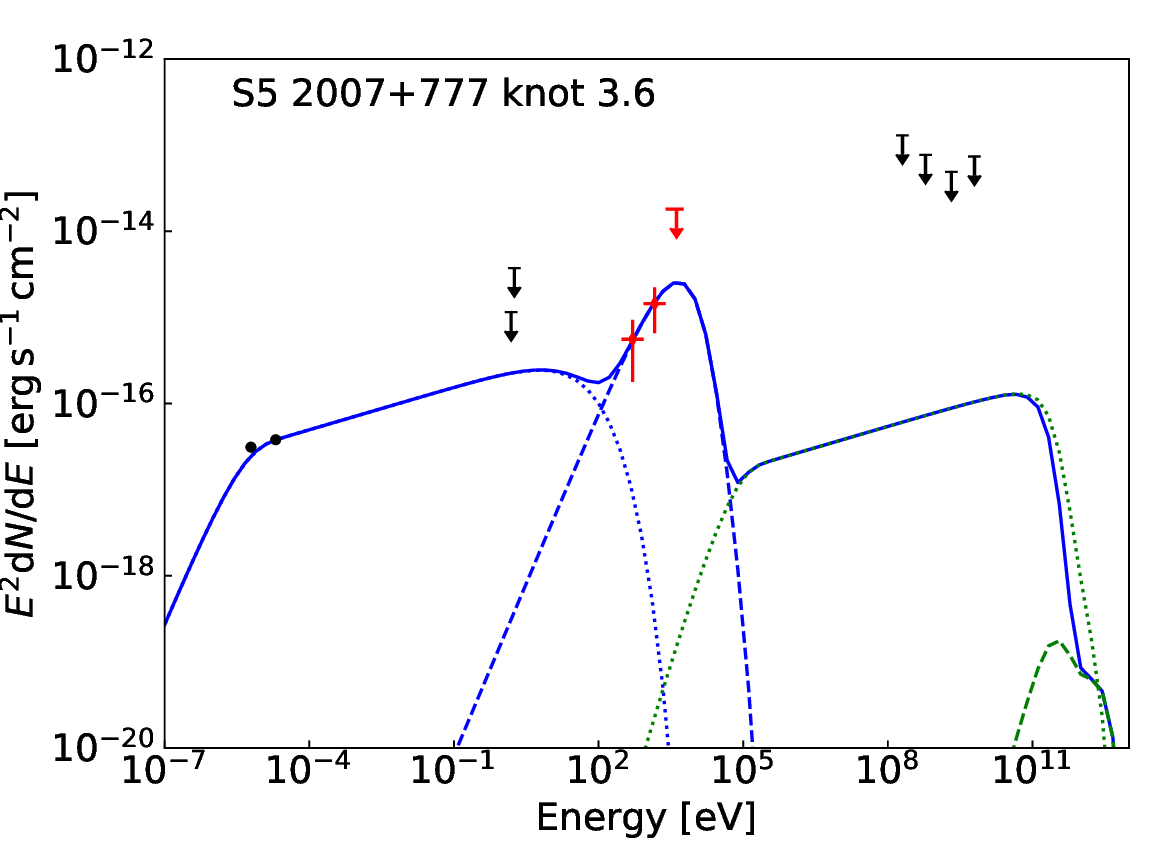}
\includegraphics[scale=0.29]{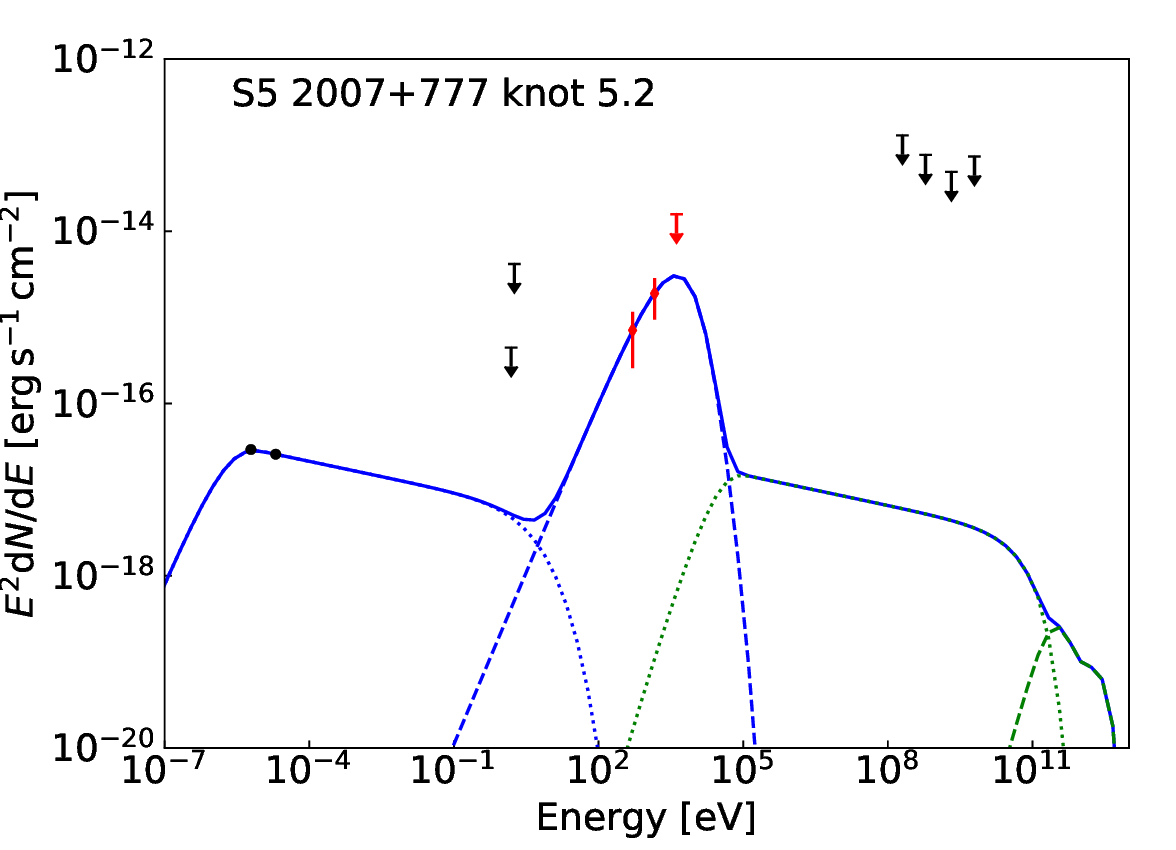}
\includegraphics[scale=0.29]{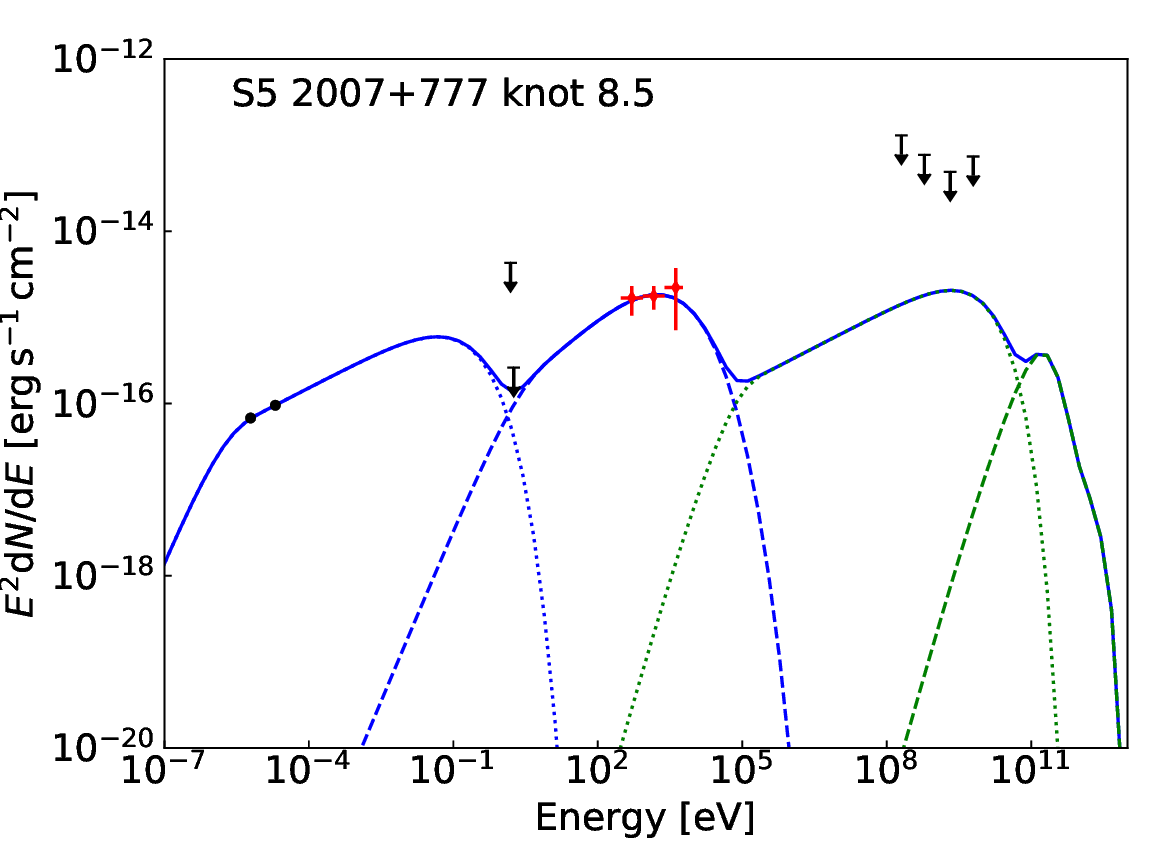}
\includegraphics[scale=0.29]{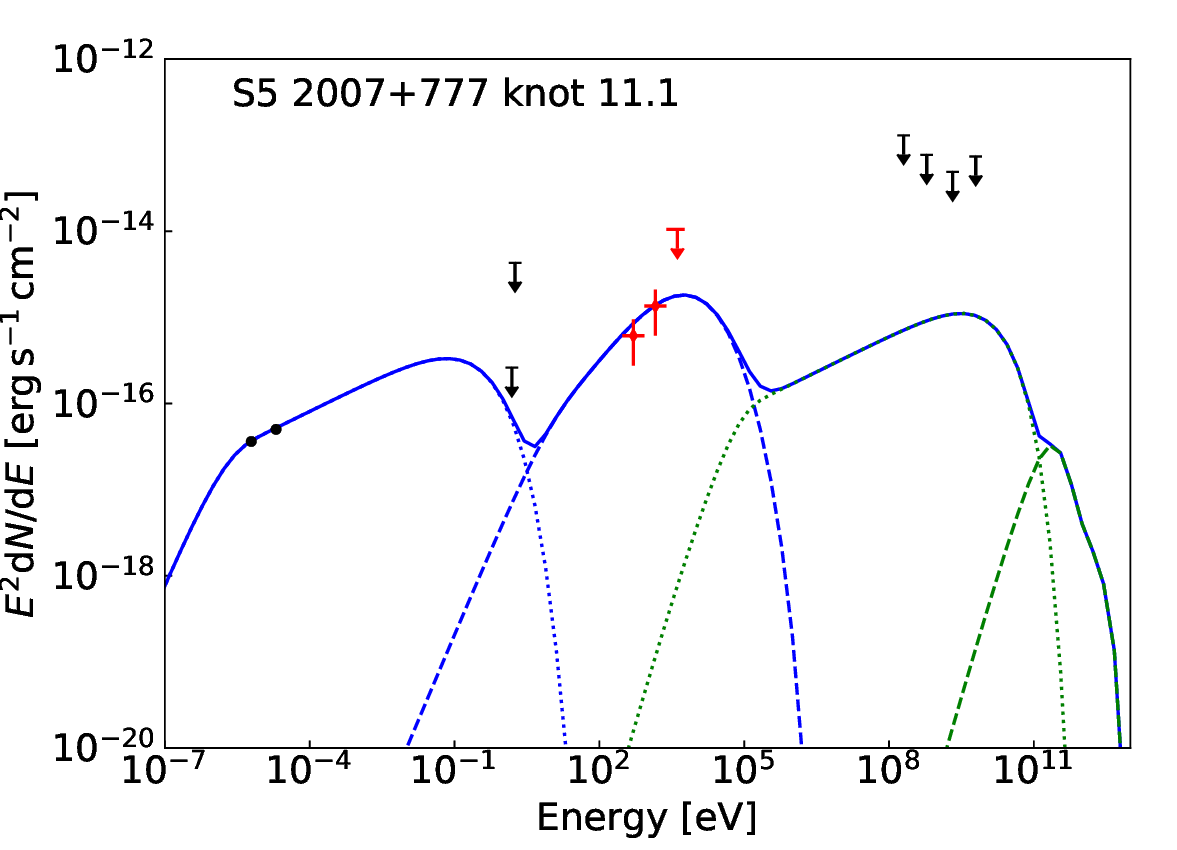}
\includegraphics[scale=0.29]{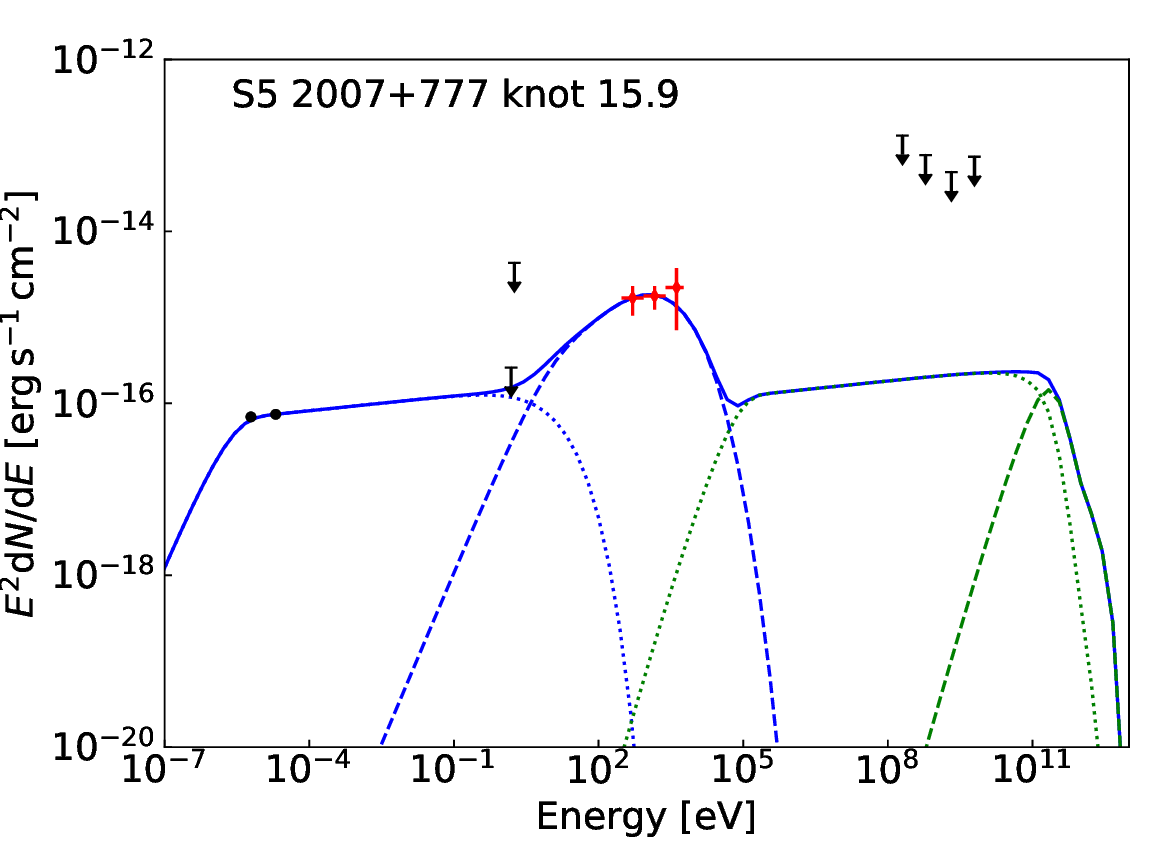}
\caption {Fitting results and broad-band SEDs of the \xray knots in 3C 353 and S5 2007+777.
The styles of data and lines are the 
same as in Figure~\ref{fig:1}.}
\label{fig:3}
\end{figure*}

We apply the aforementioned shear acceleration model to the multi-wavelength observations of the 24 
selected knots. The best-fit parameters and their derived parameters are listed in Table~\ref{tab:fitting} and \ref{tab:environment}.
We also show the best-fit SEDs in Figure~\ref{fig:1}, \ref{fig:2}, and \ref{fig:3}. 
In these figures, the red points or upper limits are \chandra\ or \fermi \ data that have been re-analyzed 
in this paper, the black data points are taken from the references, see Sect.~\ref{section:Chandra_data} 
for details. The lines represent the SED fitting with emission the maximum-likelihood value. 
The individual contributions by the two populations are marked with dotted and dashed lines, respectively. 

We divide these sources into three sub-groups based on their wavelength coverage in the data-set.
(1) The knots in 3C 273, shown in Figure~\ref{fig:1}, have multi-wavelength measurements with the largest data-set, which provide the tightest constraint on the model parameters.
We take the corner plot of knot D1+D2H3 for 3C 273 as an example to show the relationship between the different parameters in Figure~\ref{figure:3}, the maximum likelihood parameter vector is indicated with the cross.
(2) The knots in the sources 3C 403, 3C 17, Pictor A, 3C 111, and PKS 2152-699 have also multi-wavelength measurements but with less data points, as shown in Figure~\ref{fig:2}.
For example, there is only one radio data point for the knots of 3C 403, Pictor A, and 3C 111, and 
the error bars in the \xray data are slightly larger due to the lower photon statistics.
(3) For the knots in 3C 353 and S5 2007+777 in Figure~\ref{fig:3}, optical measurements are missing or 
only upper limits available. Thus the constraint on $E_{\rm cutoff1}$ and $E_{\rm min2}$ is weak.

From Table~\ref{tab:fitting}, we find that the parameters change only slightly for different knots 
in the same jet, especially for 3C 273, which contains plenty of data points.
In general, the lower-energy electron population has a higher total energy content in all the sources, 
with 
$W_{\rm e,1}\sim (10^{54} - 5\times10^{58})$ ergs, while $W_{\rm e,2}\sim (10^{51} - 2\times10^{56})$ ergs 
for the high-energy electron population. 

The magnetic field strength is in the range 
$B\sim (1 - 14)~\mu $G for the different knots. 
Within the jet, the magnetic field varies slightly for the different knots. For 3C 273 and 3C 403, 
there is a slightly increasing trend for the magnetic field strength of the knots.
For the low-energy electron population, the spectral indices range from $\alpha_{\rm 1}\sim 1.8 - 3.2$, 
with a significant clustering around $2.5$, 
and the typical cutoff energies are in the range $E_{\rm cutoff1} \sim (0.4 - 26)$ TeV.

For the high-energy population, we find
$E_{\rm min2}\sim (0.4 - 55)$ TeV 
and 
$w \sim (2 - 10)$.
The difference in the shear viscous parameter ($w$) relates to their jet profiles through 
Eqs.~\ref{eq:wlinear} and \ref{eq:wpowerlaw}. 
The corresponding spectral index $\alpha_{-}$ of the high-energy electron population can 
be obtained from Eq. \ref{eq:spm}, and is in the range $\sim 1.4 - 3.2$, with some 
significant clustering around $2$.
Generally, a harder spectrum requires a higher spine velocity, as shown in
Table~\ref{tab:environment}.
For both power-law and linearly decreasing velocity profiles we find jet-spine
velocities that are mostly compatible with mildly relativistic (i.e., $\Gamma 
\lesssim 4$) flow speeds, perhaps apart from 3C 111 (K14), 3C 353 (W47), and 3C 17 (S11.3).
In general, the derived spine velocities for a power-law profile are slightly 
smaller than the ones for a linear profile. There are velocity constraints or 
measurements for some jets (such as 3C 273 and 3C 111) from their proper motions \citep{Meyer2016, Oh2015}, our derived jet velocities are generally in agreement with them. 

For different knots in the same jets, the variation of velocities ($\beta_{\rm 0, l}$ 
or $\beta_{\rm 0, p}$) is insignificant, suggesting that the X-ray jet can maintain 
its speed over a large scale.
In particular, for 3C 273, the knot speeds differ only slightly.
This is consistent with radio observations, which suggested that the jet of 3C 273 
does not decelerate substantially from knot A to knot D1 \citep{Meyer2014,Conway1993}.
For 3C 353, the model indicates that the jet speed is higher in knot W47 than in knot 
E23 and knot E88.
We note that knot W47 belongs to the counter-jet (at a distance in between the other 
two knots), while E23 and E88 belong to the main jet \citep{Kataoka2008}.

The cut-off energy of the high-energy population can be derived from Eq. \ref{eq:true_gmax}, 
and can well exceed 100 TeV (e.g., in 3C 273). In several cases, however, particularly 
for sources of sub-group~(3), e.g., 3C 353, the cut-off energy is not really constrained 
given the current data. 
A decreasing trends of $E_{\rm e,max}$ from inner to outer knots can be found 
in most knots of 3C 273 and 3C 403.


In~Table~\ref{tab:environment}, we also show the ratio between the knot power $P_{\rm knot}$ and 
the Eddington luminosity $L_{\rm edd}$ for the X-ray knots in FR II jets, where we calculate 
$P_{\rm knot}$ based on the velocity from the linear profile,
\begin{equation}
    P_{\rm knot} \simeq \frac{(W_{\rm e,1}+W_{\rm e,2}) c \beta_{\rm 0, l}}{2L_{\rm knot}/\sin \theta},
    \label{eq:luminosity}  
\end{equation} 
where $c$ is the speed of light. As we set $E_{\rm min1}\geq2.5\,{\rm GeV}>m_p c^2$, this is a good approximation of the jet kinetic energy.
Values of the viewing angle $\theta$ for different jets are discussed in Section \ref{section:Chandra_data}. 
For 3C 273, we use $\theta = 7\deg$. For 3C 353 and 3C 403, we use the lower limits on the 
viewing angles $\theta = 60\deg$ and $\theta = 45\deg$, respectively. 
For the other sources, we adopt the upper limits of the viewing angle $\theta$. 
The length of the knots $L_{\rm knot}$ is listed in Table~\ref{tab:regions}.
Typically, the resultant knot power for the different jets is 
in the range $P_{\rm knot}\sim 10^{42}-10^{46} \rm erg\ s^{-1}$.
For 3C 17, $\theta$ is unknown, hence the Doppler factor $\delta$ is uncertain, thus we assume $\delta=\Gamma$ to obtain $P_{\rm knot}$ for S11.3 and S3.7.
In general, the
required high power is essentially driven by the first electron component.

The Eddington luminosity $L_{\rm edd}=1.25\times10^{38}\, (\rm M_{\rm BH}/\msun)\, 
\rm erg\space \ s^{-1}$ can be obtained using the SMBH masses ($\rm M_{\rm BH}$) reported 
in Section \ref{section:Chandra_data}. 
We do not employ $L_{\rm edd}$ for 3C 353 and PKS 2152-699 given the lack of 
information on their SMBH mass. Instead, for PKS 2152-699, \citet{Breiding2023} 
have estimated a time-averaged jet kinetic power ($L_{\rm kin}$) $\sim 4\times 
10^{44} \rm erg\space \ s^{-1}$, our result ($\sim 1.7\times 
10^{44} \rm erg\space \ s^{-1}$) is consistent with their findings, which also take into account the thermal energy of the gas.
For all the knots, we find that the power required to reproduce the multi-wavelength 
emission is smaller than the Eddington luminosity, see Table~\ref{tab:environment}, 
i.e. $P_{\rm knot}/L_{\rm edd}\sim (10^{-4}-0.2)$.


We note that we originally did not consider beaming effects ($\delta$ = 1), except for 3C 17. However, the 
possibility of high flow speeds up to 0.99 (bulk Lorentz factors $\Gamma \sim$ 10) as 
inferred for knots HST-32 (Pictor A), K14 (3C~111), and W47 (3C~353), 
indicates that relativistic effects and Doppler boosting could be important,  
thus we obtain the Doppler factor for these knots, where $\delta^{-1}=\Gamma(1-\beta_{\rm 0, l} \cos{\theta})$ and $\Gamma = (1-\beta_{\rm 0, l}^{2})^{-\frac{1}{2}}$.


%


\section{Summary and discussion} \label{section:conclusion}

In this work we have studied a large sample of \xray knots from FR II jets within the 
framework of gradual shear acceleration and constrained the jet properties by modeling 
their multi-wavelength data.
For this, we reanalyzed \chandra \ ACIS data for 15 knots in five sources taking new observations into account, and analyzed Fermi-LAT Pass 8 data for five jets with archival data. 
The \xray spectra are compatible with a single power-law model in the energy range 0.3
-7.0 keV. 
The resultant \xray photon indices reveal variations ranging from 0.5 to  1.2.
The photon indices in the X-ray energy band are clearly different from those in the 
radio and the optical band, indicating that the emission cannot be explained by 
synchrotron radiation from a single population of electrons. Hence we explore a scenario 
where two populations of electrons contribute to the observed emission. In particular,
we consider the high-energy electron population to be energized by shear acceleration 
and being responsible for the \xrays \citep[e.g.,][]{Tavecchio2021,Wang2021MNRAS}. 

Our model for the two electron populations has seven major free parameters: the total energy 
($W_{\rm e,1}$), the spectral index ($\alpha_{\rm 1}$), and the cutoff energy ($E_{\rm 
cutoff1}$) of the low-energy electron population, the total energy ($W_{\rm e,2}$) and the 
minimum energy ($E_{\rm min2}$) of the high-energy electron population, shear viscosity 
parameter ($w$), and the mean magnetic field ($B$), as defined in Eqs. (\ref{eq:injected 
electron spectrum}-\ref{eq:wpowerlaw}). 
We use the Naima software package to perform the fitting of the multi-wavelength SEDs 
and to derive the best-fit and uncertainty distributions of those parameters through 
the MCMC algorithm. 
According to the wavelength coverage in the data-set, we have divided our sample into three subgroups, i.e., (1) the knots in 3C 273, (2) the knots in the sources 3C 403, 3C 17, Pictor A, 3C 111, and PKS 2159-699, (3) the knots in 3C 353 and S5 2007+777.  

The results are summarized in Tables \ref{tab:fitting} and \ref{tab:environment}, and 
Figures \ref{fig:1}-\ref{fig:3}.
We find that in these sources, the magnetic field is between $B\sim (1-14)~\mu$G.
For the shear-accelerated electron population, an injection at $E_{\rm min2}\sim 
(0.4-55)$ TeV is required and the cutoff energy is typically around some hundreds of 
TeV. 
The shear viscous parameter ($w$) is typically in the range of $w \sim (2,10)$, 
corresponding to electron spectral indices $\alpha_{-}$ in the range $\sim 
1.8 - 3.2$. 
With the exemption of S11.3 (3C 17) where a hard spectrum appears
and where an IC/CMB interpretation might be possible due to its uncertain 
jet inclination,
our results indicate that shear acceleration can be an efficient mechanism for 
accelerating electrons to high energy, producing the required particle spectra. 
The corresponding spine velocities are in the range $\beta_{\rm 0, l} \sim0.69 - 0.99$ for a linear profile, and $\beta_{\rm 0, p}\sim0.61 - 0.97$ for 
a power-law profile, and (with possible exception of  K14, W47, and S11.3) in principle 
all compatible with mildly relativistic ($\Gamma \lesssim 4$), large-scale jet 
flow speeds.
The small difference between the derived $\beta_{\rm 0, l}$ and $\beta_{\rm 0, p}$ 
is in agreement with the expectation that the spectral indices depend less on
velocity profiles for higher-velocity spines \citep{Rieger2022a,Wang2023MNRAS}. 
Within the jet, the derived velocities for different knots are statistically 
consistent with each other. 
For all the knots, we find that the required power to produce the multi-wavelength emission is smaller than the Eddington luminosity with a ratio $P_{\rm knot}/L_{\rm edd}\sim (10^{-4}-0.2)$. 

The parameters of the knots in 3C 273 can be tightly constrained, which allows the study of possible variations in the knots.
No significant variations are found for the parameters of the low-energy 
population ($E_{\rm cutoff1}$ and $\alpha_{\rm 1}$), while some variations are found for the parameters 
($E_{\rm e,max}$) of the high-energy population, especially in C2 and D1+D2H3. 
These differences further support that the two electron populations are produced by different processes. 
We also found that except for C2, there is a decreasing tendency for $\alpha_{\rm X}$ and a 
decreasing trend for $E_{\rm e,max}$ from the inner to the outer knots, while the derived velocities 
are compatible with each other. 
The change of the magnetic field may be related to the dynamics of the jet, which can affect shear acceleration via the changing of velocity profiles or the particle injection process. 
This needs to be explored in the future. 

The jets of AGNs are potential ultra-high-energy cosmic-ray (UHECR) accelerators according to 
the Hillas criterion \citep{Hillas1984, Aharonian2002MNRAS}. 
In the framework of shear acceleration, it is found that the maximum energies UHECRs may reach  
is $E_{\rm p,max} \simeq 3Z\xi^{\frac{1}{2 - q}}\Delta W_{\rm sh,0.1}\left ( \frac{B}{30\mu \rm G} \right ) 
\rm EeV$ \citep{Rieger2019}, where $\xi\leq1$ is the turbulence energy density ratio and $Z$ is 
the atomic number. 
Hence, in the case of strong turbulence with $\xi=1$, protons and nuclei could in principle be 
accelerated to $E_{\rm p,max}\sim (1 - 30)Z$ EeV
in those FR II sources through shear acceleration. This provides further support that the 
large-scale jets of FR II radio galaxies could serve as UHECR accelerators. 

The current analysis substantiates a picture where the X-ray emission from large-scale 
AGN jets is predominantly related to synchrotron radiation of a second population of 
electrons reaching multi-TeV energies. 
Meanwhile, deep multi-wavelength observations of FR II jets have revealed a general trend that the X-ray emission region is narrower than the radio one (e.g. \citet{Marchenko2017}) and displays an offset with the radio along the jet (e.g. \citet{Kataoka2008}). 
In the framework of shear acceleration, this may be related to the shearing profile of the jet, where the velocity gradient may be nonuniform in the jet. 
For example, in the outer sheath the velocity gradient can be smaller than at the interface of the spine and sheath as indicated by the simulations \citep{Wang2023MNRAS}, thus the particle acceleration in the outer sheath may be less efficient.
Such details can be investigated by high-resolution simulations of full jet propagation in the future. 

\section*{acknowledgements}   \label{acknowledgements}
This work is supported by the National Natural Science Foundation of China (Grant No.12133003, 12103011, U1731239, and U2031105), Guangxi Science Foundation (grant No. AD21220075). 
J.S.W. acknowledges the support from the Alexander von Humboldt Foundation. 
FMR acknowledges support by the German Science Foundation under DFG RI 1187/8-1.


\section{data availability} \label{data availability}
The \fermi \ data used in this work are publicly available, which is provided online by the NASA-GSFC Fermi Science Support Center\footnote{\url{https://fermi.gsfc.nasa.gov/cgi-bin/ssc/LAT/LATDataQuery.cgi}}. 
The \chandra\ ACIS data used in work are publicly available, which is provided online by the Chandra Data Archive\footnote{\url{https://cda.harvard.edu/chaser/mainEntry.do}}. 

\bibliographystyle{mnras}
\bibliography{ms}

\appendix 
\section{Figures} \label{appendix_figure}
\clearpage
\begin{figure}
	\centering
    \includegraphics[height=12cm,width=18cm]{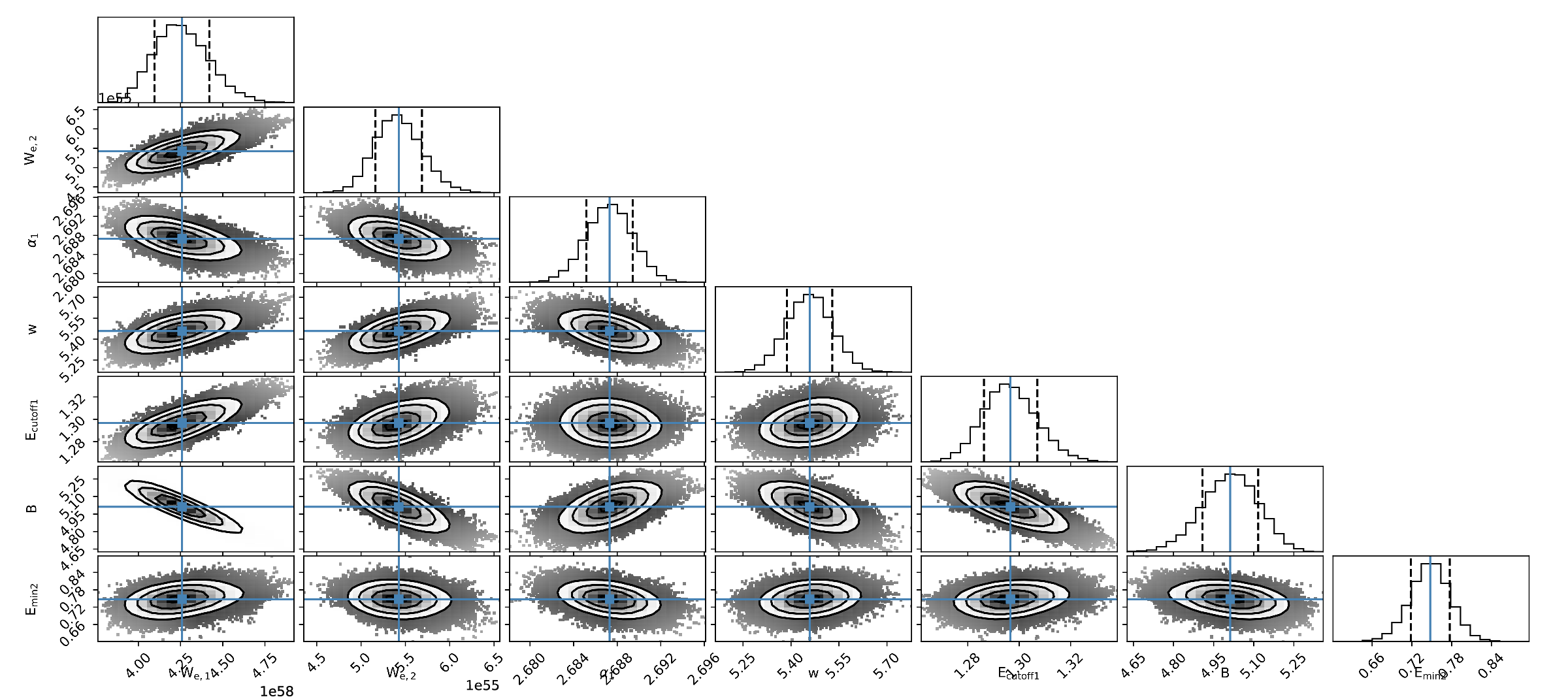}
	\caption{\textbf{The corner image of knot D1+D2H3 of 3C 273.}}
	\label{figure:3}
\end{figure}

\section{Tables}\label{appendix_table}
\begin{table*}
\centering
\caption{$Chandra$ observations of the re-analyzed FR II sources.}
\begin{tabular}{cccccccccccc}
\hline
\hline
Source&${\rm ObsID}$&${\rm Exp Time}$[ks]&${\rm Start Date}$ (YYY-MM-DD)&Source&${\rm ObsID}$&${\rm Exp Time}$[ks]&${\rm Start Date}$ (YYY-MM-DD)\\ \hline \hline
\textit3C 111&$14990$&$92.100$&$2013-01-10$ & 3C 273&$459$&$38.670$&$2000-01-10$\\
 &$16219$&$143.41$&$2014-11-04$ & &$1711$&$27.120$&$2000-06-14$\\
 &$19615$&$22.540$&$2017-12-26$ & &$1712$&$27.450$&$2000-06-14$\\
 &$19616$&$23.500$&$2019-01-03$ & &$2463$&$26.690$&$2001-06-13$\\
 &$20907$&$28.160$&$2017-12-29$ & &$2464$&$29.460$&$2001-06-13$\\
 &$20908$&$27.230$&$2017-12-29$ & &$2471$&$24.890$&$2001-06-15$\\
 &$22023$&$15.690$&$2018-12-30$ & &$3456$&$24.530$&$2002-06-05$\\
 &$22024$&$19.600$&$2018-12-29$ & &$3457$&$24.850$&$2002-06-05$\\
 &$22025$&$16.180$&$2019-01-06$ & &$3574$&$29.340$&$2002-06-04$\\
 &$22026$&$12.760$&$2019-01-07$ & &$4430$&$27.150$&$2003-07-07$\\
3C 403&$2968$&$49.470$&$2002-12-07$ & &$4431$&$26.420$&$2003-07-07$\\
 &$12741$&$7.9500$&$2010-11-27$ & &$4876$&$37.460$&$2003-11-24$\\
PKS 2152-699&$11530$&$56.750$&$2010-01-22$ & &$4877$&$34.860$&$2004-02-10$\\
 &$12088$&$58.360$&$2010-01-20$ & &$4878$&$34.090$&$2004-04-26$\\
 &$16083$&$121.17$&$2014-07-19$ & &$4879$&$35.580$&$2004-07-28$\\
 &$16084$&$57.310$&$2014-07-16$ & &$5169$&$29.680$&$2004-06-30$\\
S5 2007+777&$5709$&$36.050$&$2005-05-23$  & &$5170$&$28.400$&$2004-06-30$\\
 &   &   &   &   &$7364$&$2.0100$&$2007-01-15$\\
 &   &   &   &   &$7365$&$2.1200$&$2007-07-10$\\
 &   &   &   &   &$8375$&$29.550$&$2007-06-25$\\
 &   &   &   &   &$9703$&$29.700$&$2008-05-08$\\
 &   &   &   &   &$14455$&$29.550$&$2012-07-16$\\
 &   &   &   &   &$17393$&$29.540$&$2015-07-14$\\
 &   &   &   &   &$18421$&$29.600$&$2016-06-27$\\
 &   &   &   &   &$19867$&$26.910$&$2017-06-26$\\
 &   &   &   &   &$20709$&$29.570$&$2018-07-04$\\
 &   &   &   &   &$21815$&$29.590$&$2019-07-03$\\
 &   &   &   &   &$22828$&$28.410$&$2020-07-06$\\
 &   &   &   &   &$24585$&$25.590$&$2021-06-10$\\
 \hline
 \end{tabular}
\label{tab:1}
\end{table*}

\begin{table*}
\centering
\caption{\xray positions and sizes of the knots}
\begin{threeparttable}
\begin{tabular}{ccccccccc}
\hline
\hline
Source&Knot&${\rm RA (hh:mm:ss)}$&${\rm Dec (dd:mm:ss)}$&${L_{\rm knot} (\rm \arcsec)}$\tnote{b} &${ W_{\rm knot} (\rm \arcsec)}$\tnote{c} \\\hline \hline
\textit 3C 273 \tnote{a} &A&$12:29:06.14$&$+02:02:58.87$&0.94 (2.53 $\rm kpc)$&1.33 (3.60 $\rm kpc)$\\
 &B1+B2&$12:29:06.03$&$+02:02:57.55$&0.94 (2.53 $\rm kpc)$&1.31 (3.53 $\rm kpc)$\\
 &B3+C1&$12:29:05.94$&$+02:02:56.13$&0.77 (2.08 $\rm kpc)$&1.05 (2.84 $\rm kpc)$\\
 &C2&$12:29:05.88$&$+02:02:55.11$&0.70 (1.89 $\rm kpc)$&1.02 (2.76 $\rm kpc)$\\
 &D1+D2H3&$12:29:05.81$&$+02:02:53.62$&1.01 (2.72 $\rm kpc)$&1.02 (2.74 $\rm kpc)$\\
3C 403&F1&$19:52:19.12$&$+02:30:33.30$&3.60 (4.06 $\rm kpc)$&3.60 (4.06 $\rm kpc)$\\
 &F6&$19:52:17.57$&$+02:30:33.20$&3.60 (4.06 $\rm kpc)$&3.60 (4.06 $\rm kpc)$\\
3C 17&S3.7&$00:38:20.77$&$-02:07:41.60$&0.46 (1.60 $\rm kpc)$&0.18 (0.60 $\rm kpc)$\\
 &S11.3&$00:38:21.20$&$-02:07:45.90$& 0.40 (1.40 $\rm kpc)$&0.30 (1.00 $\rm kpc)$\\ 
Pictor A&HST-32&$05:19:46.69$&$-45:46:37.35$&3.60 (2.48 $\rm kpc)$&3.60 (2.48 $\rm kpc)$\\
 &HST-106&$05:19:39.73$&$-45:46:22.80$&3.60 (2.48 $\rm kpc)$&3.60 (2.48 $\rm kpc)$\\
 &HST-112&$05:19:39.20$&$-45:46:21.80$& 4.00 (2.80 $\rm kpc)$ &2.00 (1.40 $\rm kpc)$\\ 
3C 111 &K14&$04:18:22.50$&$+38:01:43.06$&3.60 (3.42 $\rm kpc$)&3.60 (3.42 $\rm kpc$)\\
 &K30&$04:18:23.48$&$+38:01:50.60$&2.49 (2.42 $\rm kpc)$&1.55 (1.50 $\rm kpc)$\\
 &K61&$04:18:25.66$&$+38:02:05.26$&2.89 (2.80 $\rm kpc)$&2.13 (2.07 $\rm kpc)$\\
PKS 2152-699&D&$21:57:07.07$&$-69:41:14.77$&1.73 (0.97 $\rm kpc)$&1.73 (0.97 $\rm kpc)$\\
3C 353 &E23&$17:20:29.70$&$-00:58:41.60$&1.20 (0.72 $\rm kpc$)&1.20 (0.72 $\rm kpc$)\\
 &E88&$17:20:32.80$&$-00:58:26.60$&1.50 (0.90 $\rm kpc$)&1.50 (0.90 $\rm kpc$)\\
 &W47&$17:20:29.70$&$-00:58:41.60$&2.00 (1.20 $\rm kpc$)&2.00 (1.20 $\rm kpc$)\\
S5 2007+777&K3.6&$20:05:30.25$&$+77:52:42.31$&0.83 (3.98 $\rm kpc)$&0.83 (3.98 $\rm kpc)$\\
 &K5.2&$20:05:29.69$&$+77:52:42.50$&0.62 (3.00 $\rm kpc)$&0.62 (3.00 $\rm kpc)$\\
 &K8.5&$20:05:28.45$&$+77:52:41.18$&0.62 (3.00 $\rm kpc)$&0.62 (3.00 $\rm kpc)$\\
 &K11.1&$20:05:27.51$&$+77:52:39.39$&0.62 (3.00 $\rm kpc)$&0.62 (3.00 $\rm kpc)$\\
 &K15.9&$20:05:26.29$&$+77:52:38.66$&0.62 (3.00 $\rm kpc)$&0.62 (3.00 $\rm kpc)$\\
\hline
\end{tabular}

\begin{tablenotes}
\footnotesize
\item[a] As the adjacent knots in 3C 273 jet cannot be resolved in the \xray band, we combined multiple knot regions to perform the spectral analysis. 
\item[b]$L_{\rm knot}$ is the half width at half maximum along the jet.
\item[c]$W_{\rm knot}$ is the half width at half maximum transverse to the jet.
The $L_{\rm knot}$ and $W_{\rm knot}$ of 3C 273, 3C 17, 3C 353, and Pictor A are taken from \citet{Jester2006}, \citet{Massaro2009}, \citet{Kataoka2008}, and \citet{Gentry2015}, respectively. 
\end{tablenotes}
\end{threeparttable}
\label{tab:regions}
\end{table*}

\bsp

\label{lastpage}
\end{document}